%% file: all.tex
\documentclass
[paper=A4,abstract=true,numbers=endperiod,toc=bibliography]
{scrartcl}
\usepackage{amsmath,amssymb}
\usepackage{graphicx}
\input{title}
\begin{document}
\maketitle
\input{abstract}
\tableofcontents
\input{section1}

\input{section2}

\input{section3}

\input{section4}
\input{section5}
\input{section6}

\input{section7}
\input{appendix}

\input{bib}
\end{document}

%% file: title.tex
\title{Internal bores and gravity currents in a two-fluid system}
\author{Kiyoshi Maruyama\\
                \footnotesize Department of Earth and Ocean Sciences,
                \footnotesize National Defense Academy,\\
                \footnotesize Yokosuka, Kanagawa 239-8686, Japan}
\date{\footnotesize \today}
\makeatletter
    
    \@addtoreset{equation}{section}
\makeatother

%% file: abstract.tex
\begin{abstract}
In this paper, a unified theory of internal bores and 
gravity currents is presented within the framework of 
the one-dimensional two-layer shallow-water equations. 

The equations represent four basic physical laws: 
the theory is developed on the basis of these laws. 
Though the first three of the four basic laws are apparent, 
the forth basic law has been uncertain. 
This paper shows first that this forth basic law 
can be deduced from the law which is called 
in this paper the conservation law of circulation. 

It is then demonstrated that, 
within the framework of the equations, 
an internal bore is represented by a shock 
satisfying the shock conditions 
that follow from the four basic laws. 
A gravity current can also be treated 
within the framework of the equations if the front conditions, 
i.e.\ the boundary conditions to be imposed 
at the front of the current, are known. 
Basically, the front conditions for a gravity current 
also follow from the four basic laws. 
When the gravity current 
is advancing along a no-slip boundary, however, 
it is necessary to take into account the influence 
of the thin boundary layer formed on the boundary; 
this paper describes how this influence can be evaluated. 

It is verified that the theory 
can satisfactorily explain the behaviour of internal bores 
advancing into two stationary layers of fluid. 
The theory also provides a formula for the rate of advance 
of a gravity current along a no-slip lower boundary; 
this formula proves to be consistent with some empirical formulae. 
In addition, some well-known theoretical formulae 
on gravity currents turn out to be obtainable 
on the basis of the theory. 
\end{abstract}

%% file: section1.tex
\section{Introduction}
The one-dimensional two-layer shallow-water equations 
are often used to describe the motion of 
two superposed fluids of different density in a channel; 
since the equations are mathematically tractable, 
the use of them is quite advantageous. 
Within the framework of these equations, 
an internal bore is represented by a `shock', 
i.e.\ a discontinuity that divides two continuous solutions. 
The representation enables us to deal with an internal bore 
without detailed information on its structure. 
On the other hand, certain boundary conditions are necessary 
at the position of such a discontinuity to connect 
the solutions on the two sides of the discontinuity. 
Our first aim in the present study is to formulate 
these boundary conditions, i.e.\ the `shock conditions', 
appropriate to the equations. 

The one-dimensional two-layer shallow-water equations 
form a system of four partial differential equations; 
this implies that the equations represent four basic physical laws. 
The shock conditions for the equations, 
as well as the equations themselves, 
are derived from these laws (see e.g.\ Whitham 1974, \S\,5.8). 
The first three of the four basic laws are apparent: 
the conservation laws of mass for the upper layer, 
of mass for the lower layer, and of momentum for the layers together. 
Surprisingly, however, 
the fourth basic law of the equations is still uncertain. 
Because of this uncertainty, 
the shock conditions for the equations 
are not completely determined yet. 

The shock conditions for the one-dimensional two-layer 
shallow-water equations were first studied by Yih \& Guha (1955). 
They employed the conservation law of momentum for the lower layer, 
or equivalently the same conservation law for the upper layer, 
as the fourth basic law of the equations. 
To derive the shock conditions from their set of basic laws, 
however, it is necessary to evaluate 
the exchange of momentum between the layers 
which occurs inside an internal bore. 
By approximating this exchange of momentum on some assumptions, 
they obtained a set of shock conditions. 

Chu \& Baddour (1977) and Wood \& Simpson (1984) also suggested 
two other sets of shock conditions. 
One of the sets is obtainable using 
the conservation law of mechanical energy for the upper layer 
as the fourth basic law of the equations, and the other 
using the same conservation law for the lower layer. 
Wood \& Simpson conducted, in addition, 
experiments on internal bores advancing into 
two stationary layers of fluid with a small density difference. 
They demonstrated that the former set of shock conditions, 
and also Yih \& Guha's, can account for the experimental results 
so long as the amplitudes of the bores are small enough. 
On the other hand, Klemp, Rotunno \& Skamarock (1997) later showed 
that the experimental results of Wood \& Simpson 
can be explained, irrespective of the amplitudes of the bores, 
by the latter set of shock conditions. 
However, the latter set of shock conditions 
leads us to the strange conclusion 
that no shocks can exist when the density 
in the upper layer is much smaller than that in the lower layer. 
It is evident that this conclusion contradicts 
the theory of bores in classical hydraulics. 

In the present study, special attention is paid to the law 
on the balance of circulation whose mathematical expression 
is given, for example, by Pedlosky (1987, \S\,2.2); 
this law may be called the `conservation law' of circulation 
because Kelvin's circulation theorem is deduced as its corollary. 
An essential assertion of the present study is 
that the fourth basic law of the one-dimensional two-layer 
shallow-water equations can be derived from the conservation law. 
The shock conditions obtained from the resulting 
set of basic laws can satisfactorily account for 
the experimental results of Wood \& Simpson, 
and are also consistent with the theory of bores 
in classical hydraulics. 

The one-dimensional two-layer shallow-water equations can also 
be used to study the behaviour of a gravity current in a channel, 
as proposed by Rottman and Simpson (1983) and Klemp, Rotunno \& Skamarock (1994). 
In such a study, the fluid motion behind the front 
of a gravity current is assumed to be governed by the equations. 
Accordingly, it is necessary 
to impose appropriate boundary conditions 
at the front of the gravity current 
in order to solve the equations behind the front. 
Our second aim in the present study 
is to formulate these `front conditions' 
for some important kinds of gravity currents. 

Basically, we can derive the front conditions for a gravity current 
again from the four basic laws stated above. 
However, when the gravity current 
is advancing along a no-slip boundary, 
we must include in the conditions the influence 
of the boundary layer formed on the boundary, 
no matter how thin this boundary layer may be. 
In the present study, this is done with the aid of 
the observations of gravity currents by Simpson (1972). 

Once the front conditions for a gravity current are obtained, 
a formula that gives the rate of advance of the gravity current 
as a function of its depth can be derived from the conditions. 
The determination of formulae of this kind has been one of 
the chief aims of theories of gravity currents, 
and some formulae have been proposed up to the present. 

A formula of this kind was first obtained by von K\'arm\'an (1940) 
for a gravity current advancing along a lower boundary 
into a much deeper fluid. 
We can show in the present study that, 
while his argument leading to the formula 
is unacceptable (Benjamin 1968), 
the formula itself applies 
if the fluids are allowed to slip at the lower boundary. 

About von K\'arm\'an's formula, 
Rotunno, Klemp \& Weisman (1988) later showed that, 
when the Boussinesq approximation is adequate, 
it can be derived solely on the basis of 
the conservation laws of mass and of circulation. 
The present study also supports this. 

On the other hand, Benjamin (1968) found a formula 
for a gravity current advancing along an upper boundary 
into a much heavier fluid. 
It is confirmed in the present study that 
the same formula is obtained 
for this specific kind of gravity current. 

Benjamin also argued that his formula would apply 
to other kinds of gravity currents 
if the acceleration due to gravity was replaced by suitable values. 
However, the present study 
leads us to the following conclusion: 
within the framework of 
the one-dimensional two-layer shallow-water equations, 
Benjamin's formula applies only to a gravity current 
advancing along an upper boundary into a much heavier fluid. 

Thus Benjamin's formula does not apply, 
within the framework of the equations, 
to a familiar gravity current 
advancing along a no-slip lower boundary; 
in the present study, a different formula 
can be obtained for this kind of gravity current. 
This formula proves to be consistent with 
the empirical formulae reported by Yih (1965, p.\,136), 
Simpson \& Britter (1980), and Rottman \& Simpson (1983). 

The format of this paper is as follows. 
We first derive a new mathematical expression 
of the conservation law of circulation in \S\,2; 
this is because the customary expression 
of the law is inconvenient for the subsequent analysis. 
Next, we discuss the four basic laws of the one-dimensional 
two-layer shallow-water equations in \S\,3. 
After these preliminary sections, 
internal bores are considered in \S\,4, 
and gravity currents in \S\,5. 
Section 6 gives a discussion 
on the applicability of the present theory.

%% file: section2.tex
\section{Conservation law of circulation}
The conservation law of circulation is a two-dimensional 
conservation law on a surface always composed of 
the same fluid particles, i.e.\ on a material surface. 
It is customarily expressed by an equation for 
the rate of change of the circulation 
around a closed curve moving on a material surface 
with fluid particles (see e.g.\ Pedlosky 1987, \S\,2.2). 
Though this customary expression of the law is useful, 
for example, to prove that a certain kind of flow is 
irrotational, it is inconvenient for the subsequent analysis. 
We therefore derive in this section 
a new mathematical expression of the law. 

Consider a material surface 
in a three-dimensional space occupied by a fluid. 
In order to specify the positions on the surface, we set up 
a system of coordinates $(\theta^1,\theta^2)$ on the surface. 
We call the coordinates the surface coordinates, 
and consider them to be `fixed' on the surface. 
That is to say, we consider 
a point on the material surface to be fixed on the surface 
if its surface coordinates are invariable. 

Now let $\Gamma$ be a closed curve 
fixed on the material surface, i.e.\ a closed curve 
composed of fixed points on the surface. 
The circulation around $\Gamma$ is defined as usual by 
\begin{equation}
  \label{2.10}
  \oint_\Gamma\boldsymbol{u}\cdot\boldsymbol{t}\:\mathrm{d}s, 
\end{equation}
in which $\boldsymbol{u}$ is the velocity of the fluid, 
$\boldsymbol{t}$ the unit tangent vector of $\Gamma$, 
and $\mathrm{d}s$ the element of arc length of $\Gamma$. 
The required expression of the conservation law of circulation 
is given by an equation for the rate of change of this circulation. 

In order to calculate the rate of change of (\ref{2.10}), 
we now introduce a parameter $\sigma$ such that $0\leq\sigma\leq1$, 
and assume that each of the points on $\Gamma$ 
is identified by this parameter. 
Then the surface coordinates of a point on $\Gamma$ 
can be expressed as functions of $\sigma$: 
\begin{equation}
  \label{2.20}
  \theta^\alpha=\theta^\alpha(\sigma), 
\end{equation}
where $\theta^\alpha(0)=\theta^\alpha(1)$. 
(Here and for the remaining part of this section, 
lower-case Greek indices are used to represent 
the numbers 1 and 2 for the convenience of notation; 
also, the summation convention is implied, 
i.e.\ a term in which the same index appears twice 
stands for the sum of the terms obtained by giving 
the index the values 1 and 2.) 

On the material surface, the velocity of the fluid may be regarded 
as a function of the surface coordinates and time $t$: 
\begin{equation}
  \label{2.30}
  \boldsymbol{u}=\boldsymbol{u}(\theta^\alpha,t). 
\end{equation}
Throughout this section, $\boldsymbol{u}$ is treated as such a function. 
It then follows from (\ref{2.20}) that, on $\Gamma$, 
$\boldsymbol{u}$ becomes a function of $\sigma$ and $t$. 
The position vector $\boldsymbol{R}$ of a point on the material surface 
is also a function of the surface coordinates of the point and time: 
\begin{equation}
  \label{2.40}
  \boldsymbol{R}=\boldsymbol{R}(\theta^\alpha,t). 
\end{equation}
Again, from (\ref{2.20}), the position vector of a point on $\Gamma$ 
becomes a function of $\sigma$ and $t$. 

Hence, using the parameter $\sigma$, 
we can write the circulation around $\Gamma$ as 
\begin{equation}
  \label{2.50}
  \oint_\Gamma\boldsymbol{u}\cdot\boldsymbol{t}\:\mathrm{d}s
  =\int_0^1\boldsymbol{u}\cdot\frac{\partial \boldsymbol{R}}{\partial \sigma}\:\mathrm{d}\sigma. 
\end{equation}
Its rate of change is therefore given by the following formula: 
\begin{equation}
  \label{2.60}
  \frac{\mathrm{d}}{\mathrm{d}t}\oint_\Gamma\boldsymbol{u}\cdot\boldsymbol{t}\:\mathrm{d}s
  =\int_0^1\frac{\partial \boldsymbol{u}}{\partial t}\cdot
           \frac{\partial \boldsymbol{R}}{\partial \sigma}\:\mathrm{d}\sigma 
  +\int_0^1\boldsymbol{u}\cdot\frac{\partial}{\partial \sigma}
           \left(\frac{\partial \boldsymbol{R}}{\partial t}\right)\:\mathrm{d}\sigma. 
\end{equation}

We can derive an equation for the rate of change 
of the circulation around $\Gamma$ 
from the equation of motion and (\ref{2.60}). 
To this end, however, it is necessary to introduce here 
the `surface velocity' $\boldsymbol{q}$ 
defined on the material surface by 
\begin{equation}
  \label{2.70}
  \boldsymbol{q}(\theta^\alpha,t)=\boldsymbol{u}(\theta^\alpha,t)
          -\frac{\partial\boldsymbol{R}(\theta^\alpha,t)}{\partial t}. 
\end{equation}
This is the velocity of the fluid 
which is perceived on the material surface. 
On the other hand, if the material derivative is 
denoted by $\mathrm{D}/\mathrm{D}t$, we can write 
\begin{equation}
  \label{2.80}
  \boldsymbol{u}=\frac{\mathrm{D}\boldsymbol{R}}{\mathrm{D}t}
          =\frac{\mathrm{D}\theta^\alpha}{\mathrm{D}t}\boldsymbol{a}_\alpha 
          +\frac{\partial\boldsymbol{R}}{\partial t}, 
\end{equation}
where $\boldsymbol{a}_\alpha=\partial\boldsymbol{R}/\partial\theta^\alpha$ 
are the covariant base vectors of the material surface. 
Thus we see that $\boldsymbol{q}$ can be expressed also in the form 
\begin{equation}
  \label{2.90}
  \boldsymbol{q}=q^\alpha\boldsymbol{a}_\alpha, 
\end{equation}
where $q^\alpha=\mathrm{D}\theta^\alpha/\mathrm{D}t$ are called 
the contravariant components of $\boldsymbol{q}$. 
It is evident from this expression that $\boldsymbol{q}$ 
is tangent to the material surface. 

Now, on the material surface, the equation of motion takes the form 
\begin{equation}
  \label{2.100}
  \frac{\partial \boldsymbol{u}}{\partial t}+q^\alpha\frac{\partial \boldsymbol{u}}{\partial \theta^\alpha} 
       =-\frac{1}{\rho}\nabla p+\boldsymbol{f}+\frac{1}{\rho}\boldsymbol{F}, 
\end{equation}
in which $\rho$ is the density of the fluid, $p$ the pressure, 
$\boldsymbol{f}$ the external force per unit mass, 
and $\boldsymbol{F}$ the viscous force per unit volume. 
It can readily be verified, however, that 
\begin{equation}
  \label{2.110}
  q^\alpha\frac{\partial \boldsymbol{u}}{\partial \theta^\alpha}
     =-\boldsymbol{q}\times\left(\boldsymbol{a}^\alpha\times
            \frac{\partial \boldsymbol{u}}{\partial \theta^\alpha}\right)
      +\left(\boldsymbol{q}\cdot\frac{\partial \boldsymbol{u}}{\partial \theta^\alpha}\right)
            \boldsymbol{a}^\alpha. 
\end{equation}
Here $\boldsymbol{a}^\alpha$ are 
the contravariant base vectors of the material surface, 
which are tangent to the surface and are connected with 
the covariant base vectors by 
$\boldsymbol{a}^\alpha\cdot\boldsymbol{a}_\beta=\delta^\alpha_\beta$. 
Thus we can write the equation of motion 
on the material surface as follows: 
\begin{equation}
  \label{2.120}
  \frac{\partial \boldsymbol{u}}{\partial t}
       =\boldsymbol{q}\times\left(\boldsymbol{a}^\alpha\times
             \frac{\partial \boldsymbol{u}}{\partial \theta^\alpha}\right)
        -\left(\boldsymbol{q}\cdot\frac{\partial \boldsymbol{u}}{\partial \theta^\alpha}\right)
             \boldsymbol{a}^\alpha
        -\frac{1}{\rho}\nabla p+\boldsymbol{f}+\frac{1}{\rho}\boldsymbol{F}. 
\end{equation}

Substituting (\ref{2.120}) into the first term 
on the right-hand side of (\ref{2.60}), we have 
\begin{align}
  \label{2.130}
  \frac{\mathrm{d}}{\mathrm{d}t}\oint_\Gamma\boldsymbol{u}\cdot\boldsymbol{t}\:\mathrm{d}s
 &=
 \int_0^1\left\{\boldsymbol{q}\times\left(\boldsymbol{a}^\alpha\times
                   \frac{\partial \boldsymbol{u}}{\partial \theta^\alpha}\right)
                  -\frac{1}{\rho}\nabla p +\boldsymbol{f}
                  +\frac{1}{\rho}\boldsymbol{F}\right\}
             \cdot\frac{\partial \boldsymbol{R}}{\partial \sigma}\:\mathrm{d}\sigma 
    \notag\\ 
 &\quad
    +\int_0^1\boldsymbol{u}\cdot\frac{\partial}{\partial \sigma}
             \left(\frac{\partial \boldsymbol{R}}{\partial t}\right)\:\mathrm{d}\sigma 
    -\int_0^1\boldsymbol{q}\cdot\frac{\partial \boldsymbol{u}}{\partial \sigma}\:\mathrm{d}\sigma.
\end{align}
However, since $\boldsymbol{q}=\boldsymbol{u}-\partial \boldsymbol{R}/\partial t$, 
the last two terms of (\ref{2.130}) vanish: 
\begin{equation}
  \label{2.140}
     \int_0^1\boldsymbol{u}\cdot\frac{\partial}{\partial \sigma}
             \left(\frac{\partial \boldsymbol{R}}{\partial t}\right)\:\mathrm{d}\sigma 
    -\int_0^1\boldsymbol{q}\cdot\frac{\partial \boldsymbol{u}}{\partial \sigma}\:\mathrm{d}\sigma
    =\int_0^1\frac{\partial}{\partial \sigma}
             \left(\boldsymbol{u}\cdot\frac{\partial \boldsymbol{R}}{\partial t}
                   -\frac{\boldsymbol{u}\cdot\boldsymbol{u}}{2}
             \right)\:\mathrm{d}\sigma
    =0.  
\end{equation}
This enables us to write (\ref{2.130}) as 
\begin{align}
  \label{2.150}
  \frac{\mathrm{d}}{\mathrm{d}t}\oint_\Gamma\boldsymbol{u}\cdot\boldsymbol{t}\:\mathrm{d}s
 &=
  \oint_\Gamma\left\{\boldsymbol{q}\times\left(\boldsymbol{a}^\alpha\times
               \frac{\partial \boldsymbol{u}}{\partial \theta^\alpha}\right)\right\}
               \cdot\boldsymbol{t}\:\mathrm{d}s
    \notag\\
 &\quad
  -\oint_\Gamma\frac{1}{\rho}\nabla p\cdot\boldsymbol{t}\:\mathrm{d}s 
  +\oint_\Gamma\boldsymbol{f}\cdot\boldsymbol{t}\:\mathrm{d}s
  +\oint_\Gamma\frac{1}{\rho}\boldsymbol{F}\cdot\boldsymbol{t}\:\mathrm{d}s.
\end{align}
We have thus obtained an equation for the rate of change 
of the circulation around $\Gamma$. 

Note that (\ref{2.150}) contains 
four terms on its right-hand side. 
The second and the third of them represent 
the rate of generation of circulation due to baroclinicity 
and that due to the external force respectively; 
the fourth of them may be taken 
to represent the rate of diffusion 
of circulation across $\Gamma$ (see e.g.\ Pedlosky 1987, \S\,2.2). 
However, the meaning of the first 
of the four terms is not apparent as it stands. 
Thus it is desirable to rewrite this term in a more 
physically meaningful form. 

Let $\boldsymbol{n}$ be the unit normal to the material surface, 
and let $\boldsymbol{\nu}$ be the unit vector defined on $\Gamma$ 
by $\boldsymbol{\nu}=\boldsymbol{t}\times\boldsymbol{n}$: 
the vector $\boldsymbol{\nu}$ is the unit outward normal to $\Gamma$ 
in the material surface. 
Using these vectors, we can rewrite the first term 
on the right-hand side of (\ref{2.150}) as 
\begin{equation}
  \label{2.160}
  -\oint_\Gamma\left\{\left(\boldsymbol{a}^\alpha\times
                         \frac{\partial \boldsymbol{u}}{\partial \theta^\alpha}\right)
                         \cdot\boldsymbol{n}\right\}\boldsymbol{q}\cdot\boldsymbol{\nu}\:\mathrm{d}s.
\end{equation}
However, it can easily be shown that 
\begin{equation}
  \label{2.170}
  \left(\boldsymbol{a}^\alpha\times
                \frac{\partial \boldsymbol{u}}{\partial \theta^\alpha}\right)\cdot\boldsymbol{n}
  =\omega_n,
\end{equation}
where $\omega_n=\boldsymbol{\omega}\cdot\boldsymbol{n}$ is the component 
of the vorticity $\boldsymbol{\omega}$ normal to the material surface. 
As a result, we can further rewrite (\ref{2.160}) in the form 
\begin{equation}
  \label{2.180}
  -\oint_\Gamma\omega_n\boldsymbol{q}\cdot\boldsymbol{\nu}\:\mathrm{d}s.
\end{equation}

The interpretation of (\ref{2.180}) follows 
immediately from Stokes' theorem: 
\begin{equation}
  \label{2.190}
  \oint_\Gamma\boldsymbol{u}\cdot\boldsymbol{t}\:\mathrm{d}s
 =\int\!\!\!\int_\Sigma\omega_n\:\mathrm{d}S,
\end{equation}
in which $\Sigma$ is the part of the material surface 
enclosed by $\Gamma$, and $\mathrm{d}S$ the element of area 
of the material surface. 
From (\ref{2.190}), we see that $\omega_n$ is 
the surface density of circulation, i.e.\ 
the circulation per unit area, on the material surface. 
Hence it is evident that (\ref{2.180}) represents 
the rate of advection of circulation 
across $\Gamma$ into the area $\Sigma$. 
(We use, in this paper, the term `advection' 
to refer to `transport due to fluid motion'.)

We have now found that (\ref{2.150}) can be expressed 
in the following form: 
\begin{equation}
  \label{2.200}
  \frac{\mathrm{d}}{\mathrm{d}t}\oint_\Gamma\boldsymbol{u}\cdot\boldsymbol{t}\:\mathrm{d}s
 =-\oint_\Gamma\omega_n\boldsymbol{q}\cdot\boldsymbol{\nu}\:\mathrm{d}s
  -\oint_\Gamma\frac{1}{\rho}\nabla p\cdot\boldsymbol{t}\:\mathrm{d}s 
  +\oint_\Gamma\boldsymbol{f}\cdot\boldsymbol{t}\:\mathrm{d}s
  +\oint_\Gamma\frac{1}{\rho}\boldsymbol{F}\cdot\boldsymbol{t}\:\mathrm{d}s.
\end{equation}
This is the required expression 
of the conservation law of circulation. 
It shows that the change in the circulation 
around a closed curve fixed on a material surface is caused by 
(i) the advection of circulation across the closed curve, 
(ii) the generation of circulation due to baroclinicity, 
(iii) the generation of circulation due to the external force, and 
(iv) the diffusion of circulation across the closed curve. 

The customary expression of the conservation law of circulation 
can be obtained from (\ref{2.200}). 
To demonstrate this, we now consider a closed curve $\Gamma_t$ which 
always consists of the same fluid particles on a material surface. 
From Stokes' theorem, the rate of change of the circulation 
around $\Gamma_t$ can be written as 
\begin{equation}
  \label{2.210}
  \frac{\mathrm{d}}{\mathrm{d}t}\oint_{\Gamma_t}\boldsymbol{u}\cdot\boldsymbol{t}\:\mathrm{d}s 
 =\frac{\mathrm{d}}{\mathrm{d}t}\int\!\!\!\int_{\Sigma_t}\omega_n\:\mathrm{d}S,
\end{equation}
where $\Sigma_t$ denotes the part 
of the material surface enclosed by $\Gamma_t$. 
On the other hand, 
at an arbitrary instant $t=t_0$, 
there exists a closed curve fixed on the material surface 
that coincides with $\Gamma_t$. 
We may take this closed curve 
as $\Gamma$ which appears in (\ref{2.200}). 
The rate of change of the circulation 
around this closed curve can also be written as 
\begin{equation}
  \label{2.220}
  \frac{\mathrm{d}}{\mathrm{d}t}\oint_\Gamma\boldsymbol{u}\cdot\boldsymbol{t}\:\mathrm{d}s
  =\frac{\mathrm{d}}{\mathrm{d}t}\int\!\!\!\int_\Sigma\omega_n\:\mathrm{d}S.
\end{equation}
However, it can be verified (see e.g.\ Aris 1962, \S\,10.12) 
that, at $t=t_0$, 
\begin{equation}
  \label{2.230}
  \frac{\mathrm{d}}{\mathrm{d}t}\int\!\!\!\int_{\Sigma_t}\omega_n\:\mathrm{d}S
 =\frac{\mathrm{d}}{\mathrm{d}t}\int\!\!\!\int_\Sigma\omega_n\:\mathrm{d}S
 +\oint_\Gamma\omega_n\boldsymbol{q}\cdot\boldsymbol{\nu}\:\mathrm{d}s.
\end{equation}
Thus it follows from (\ref{2.210}) and (\ref{2.220}) 
that, at $t=t_0$, 
\begin{equation}
  \label{2.240}
  \frac{\mathrm{d}}{\mathrm{d}t}\oint_{\Gamma_t}\boldsymbol{u}\cdot\boldsymbol{t}\:\mathrm{d}s
 =\frac{\mathrm{d}}{\mathrm{d}t}\oint_\Gamma\boldsymbol{u}\cdot\boldsymbol{t}\:\mathrm{d}s
 +\oint_\Gamma\omega_n\boldsymbol{q}\cdot\boldsymbol{\nu}\:\mathrm{d}s.
\end{equation}
We now substitute (\ref{2.200}) into (\ref{2.240}) 
to find that, at $t=t_0$, 
\begin{equation}
  \label{2.250}
  \frac{\mathrm{d}}{\mathrm{d}t}\oint_{\Gamma_t}\boldsymbol{u}\cdot\boldsymbol{t}\:\mathrm{d}s 
 =-\oint_{\Gamma_t}\frac{1}{\rho}\nabla p\cdot\boldsymbol{t}\:\mathrm{d}s 
  +\oint_{\Gamma_t}\boldsymbol{f}\cdot\boldsymbol{t}\:\mathrm{d}s
  +\oint_{\Gamma_t}\frac{1}{\rho}\boldsymbol{F}\cdot\boldsymbol{t}\:\mathrm{d}s.
\end{equation}
Here the path of integration on the right-hand side 
has been changed from $\Gamma$ to $\Gamma_t$ 
since they coincide at $t=t_0$. 
However, since $t_0$ is arbitrary, 
(\ref{2.250}) in fact holds for all $t$. 
This equation gives the customary expression 
of the conservation law of circulation. 

In a similar way, (\ref{2.200}) 
can conversely be derived from (\ref{2.250}). 
We conclude, therefore, that the two distinct expressions 
(\ref{2.200}) and (\ref{2.250}) are equivalent. 
Since (\ref{2.250}) describes the rate of change 
of the circulation around a closed curve 
moving on a material surface with fluid particles, 
it may be taken as the `Lagrangian' expression 
on a material surface of the conservation law of circulation. 
On the other hand, since (\ref{2.200}) describes 
the rate of change of the circulation 
around a closed curve fixed on a material surface, 
it may be taken as the `Eulerian' expression 
on a material surface of the same conservation law. 

Finally, the following fact should be noted: 
for any closed curve in a fluid, 
we can find a material surface 
on which the closed curve always exists; 
in addition, we can set up 
on the surface a system of surface coordinates 
in such a way that the surface coordinates of all the points 
on the closed curve are invariable. 
This fact implies that any closed curve in a fluid 
may be regarded as a closed curve fixed on a material surface. 
Hence it is seen that the conservation law of circulation 
is expressed about any closed curve in a fluid by (\ref{2.200}): 
the Lagrangian expression (\ref{2.250}) may 
also be considered a special form of (\ref{2.200}). 
This is the reason why 
the Eulerian expression (\ref{2.200}) has been derived. 
In the following section, (\ref{2.200}) is used in practice 
to express the conservation law of circulation.

%% file: section3.tex
\section{One-dimensional two-layer shallow-water equations}
In this section, we examine the physical implications of 
the one-dimensional two-layer shallow-water equations. 
These equations represent, as stated in \S\,1, 
four basic physical laws. 
Our first task in this section is to formulate 
these four basic laws with regard to a typical problem 
which can be handled using the equations. 
It is then demonstrated that the equations 
can really be derived from these four basic laws. 

Before starting the discussion, we introduce 
the following convention on the notation of variables: 
in this and the subsequent sections, 
dimensional variables are indicated by asterisks, 
e.g.\ dimensional time is henceforth denoted by $t_{\ast}$; 
variables without asterisks 
should be interpreted as dimensionless, 
e.g.\ dimensionless time is denoted by $t$. 
For the remainder of this paper, 
dimensionless variables are primarily used. 

\subsection{Four basic laws of the equations}
We consider the motion under gravity 
of a two-layer incompressible Newtonian fluid 
in a horizontal channel of uniform rectangular cross-section: 
the channel has a rigid upper boundary 
and is occupied entirely by the fluid. 
The width and the depth of the channel 
are denoted by $W$ and $H$ respectively. 
We assume that the two layers are divided 
by an interface of zero thickness 
whose surface tension is negligible. 
Within each of the layers, the density takes a constant value: 
$\rho_{1}$ in the lower layer and $\rho_{2}$ in the upper layer. 

In order to specify the positions in the channel, 
we construct a system of rectangular coordinates 
$(x_{\ast},y_{\ast},z_{\ast})$ by taking 
the $x_{\ast}$-axis along the centreline of the lower boundary, 
the $y_{\ast}$-axis at a right angle to the side boundaries, 
and the $z_{\ast}$-axis vertically upward. 
In this coordinate system, 
the side boundaries are expressed by $y_{\ast}={\pm}\frac{1}{2}W$, 
and the upper and lower boundaries 
by $z_{\ast}=H$ and $z_{\ast}=0$ respectively. 

Let us now suppose that the motion possesses the length scale $L$ 
which characterizes the variation in the $x_{\ast}$-direction. 
Then, to describe the motion properly, 
it is natural to use the following dimensionless coordinates: 
\begin{equation}
  \label{3.10}
  x=x_{\ast}/L,\quad y=y_{\ast}/W,\quad z=z_{\ast}/H.
\end{equation} 
We also assume that the time scale 
of the motion is $L/(\beta gH)^{\frac{1}{2}}$, 
where $\beta=(\rho_{1}-\rho_{2})/\rho_{1}$ 
and $g$ is the acceleration due to gravity. 
Then the dimensionless time $t$ defined by 
\begin{equation}
  \label{3.20}
  t=t_{\ast}(\beta gH)^{\frac{1}{2}}/L
\end{equation}
should be used together with 
the above dimensionless coordinates. 

We assume the fluid to have very small viscosity. 
Then it is reasonable to expect that thin boundary layers 
are formed on the boundaries of the channel. 
We may also assume that a free boundary layer 
surrounding the interface is formed. 
The thicknesses of these boundary layers are taken to be 
very small in comparison with the length scales $W$, $H$, and $L$. 
Then, in the system of dimensionless coordinates $(x,y,z)$, 
these boundary layers can be idealized into vortex sheets 
coincident with the boundaries and the interface: 
the influence of viscosity is confined within 
the vortex sheets, and the fluid may be regarded as effectively 
inviscid out of the vortex sheets. 
These vortex sheets are assumed never to separate 
from the boundaries and the interface. 

Suppose now that physical quantities out of the vortex sheets 
on the side boundaries are uniform in the $y$-direction. 
Then the fluid velocity out of the side vortex sheets 
must be parallel to the side boundaries. 
Thus, if the velocity scales in the $x$- and $z$-directions are 
$(\beta gH)^{\frac{1}{2}}$ and $(\beta gH)^{\frac{1}{2}}H/L$, 
we can write 
\begin{equation}
  \label{3.30}
  \boldsymbol{u}_{\ast}=(\beta gH)^{\frac{1}{2}}u(x,z,t)\boldsymbol{i}
                 +\frac{H}{L}(\beta gH)^{\frac{1}{2}}w(x,z,t)\boldsymbol{k},
                 \qquad |y|<\tfrac{1}{2},
\end{equation}
where $\boldsymbol{i}$ and $\boldsymbol{k}$ are 
the unit vectors in the $x$- and $z$-directions respectively. 

We next introduce the assumption 
that the horizontal length scale $L$ of the motion 
is very large compared with 
the depth $H$ of the channel, i.e.\ $H/L\ll 1$.
Then the vorticity out of the side vortex sheets 
can be approximated by 
\begin{equation}
  \label{3.40}
  \boldsymbol{\omega}_{\ast}
     =\frac{(\beta gH)^{\frac{1}{2}}}{H}\frac{\partial u}{\partial z} \boldsymbol{j},
     \qquad |y|<\tfrac{1}{2},
\end{equation}
where $\boldsymbol{j}$ is the unit vector in the $y$-direction. 
We assume, in addition, that the motion is irrotational 
within each of the upper and lower layers. 
It then follows from (\ref{3.40}) that 
\begin{equation}
  \label{3.50}
  u=\left\{
  \begin{array}{ll}
    u_{1}(x,t), & 0<z<h(x,t),\\[2pt]
    u_{2}(x,t), & h(x,t)<z<1.
  \end{array}
  \right.
\end{equation}
Here $z=h(x,t)$ represents the part of the interface 
out of the side vortex sheets. 

The above assumption $H/L\ll 1$ also justifies 
the hydrostatic approximation. 
Let the pressure distribution at the upper boundary 
be given by $\rho_{2}\beta gH\eta(x,t)$ for $|y|<\tfrac{1}{2}$. 
Then we can express the pressure distribution 
out of the side vortex sheets as follows: 
\begin{equation}
  \label{3.60}
  p_{\ast}=
  \left\{
  \begin{array}{lll}
    \rho_{2}\beta gH\eta +\rho_{2}gH(1-h) +\rho_{1}gH(h-z),
                         & |y|<\tfrac{1}{2}, & 0 \le z \le h,\\[2pt]
    \rho_{2}\beta gH\eta +\rho_{2}gH(1-z), 
                         & |y|<\tfrac{1}{2}, & h \le z \le 1. 
  \end{array}
  \right.
\end{equation}

Having finished the description of the problem, 
we now proceed to formulate the four basic laws 
of the one-dimensional two-layer shallow-water equations 
with regard to this problem. 
We first take up the law deducible from 
the conservation law of circulation. 

\subsubsection{Conservation law of circulation 
               for the interfacial vortex sheet}
We now focus our attention on the plane $y=0$. 
As is evident from (\ref{3.30}), 
this plane is a material surface. 
The coordinates $x$ and $z$ are employed 
as the surface coordinates for the plane, 
and the unit normal $\boldsymbol{n}$ to the plane 
is defined by $-\boldsymbol{j}$. 
It then follows that 
the surface velocity $\boldsymbol{q}_{\ast}$ on the plane 
is simply given by the velocity $\boldsymbol{u}_{\ast}$ of the fluid. 

Suppose that a closed curve $\Gamma_{0}$ 
is fixed on this plane as shown in figure 1. 
\begin{figure}[tb]
 \centering
 \includegraphics{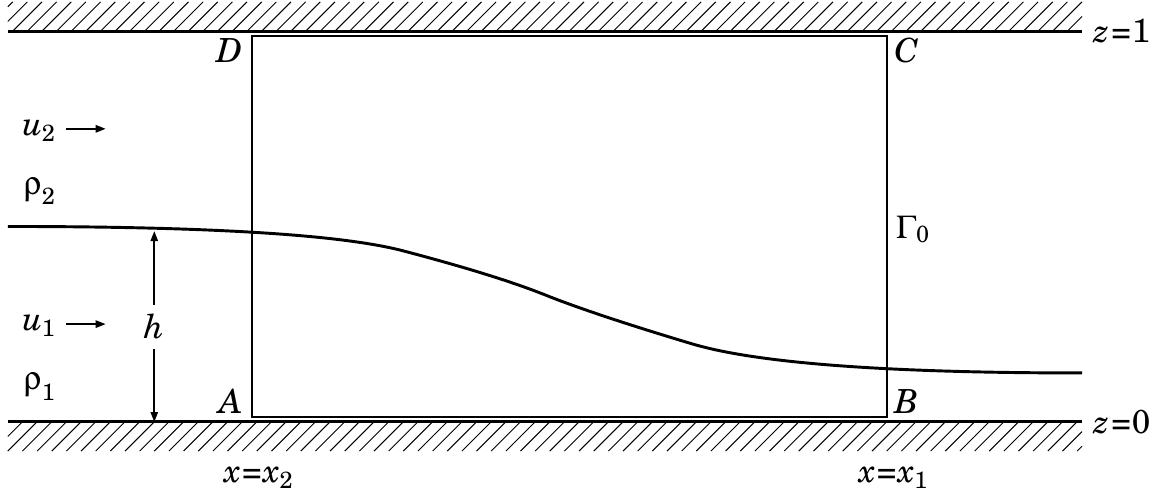}
 \caption[]{Closed curve $\Gamma_{0}$ fixed on the plane $y=0$.}
\end{figure}
This curve consists of four segments 
$AB$, $BC$, $CD$, and $DA$. 
Segment $AB$ is taken along the upper side 
of the lower vortex sheet from $x=x_{2}$ to $x=x_{1}$ 
($x_{1}$ and $x_{2}$ are 
arbitrary constants such that $x_{1}>x_{2}$), 
and $CD$ along the lower side 
of the upper vortex sheet from $x=x_{1}$ to $x=x_{2}$. 
Let us introduce here the convention that 
a zero following a plus or a minus 
denotes an infinitesimal positive number. 
Then $AB$ and $CD$ can be written as 
\[
  AB\!:\quad z=\!\mbox{}+0,\quad x_{2}\le x\le x_{1};\qquad
  CD\!:\quad z=1-0,        \quad x_{1}\ge x\ge x_{2}.
\]
On the other hand, $BC$ is taken vertically upward 
from $z=\!\mbox{}+0$ to $z=1-0$ at $x=x_{1}$, 
and $DA$ downward from $z=1-0$ to $z=\!\mbox{}+0$ at $x=x_{2}$. 
They can also be written as 
\[
  BC\!:\quad x=x_{1},\quad \!\mbox{}+0\le z\le 1-0;\qquad
  DA\!:\quad x=x_{2},\quad 1-0\ge z\ge \!\mbox{}+0.
\]

When applied to $\Gamma_{0}$, 
the conservation law of circulation is expressed 
by (\ref{2.200}) with $\Gamma$ replaced by $\Gamma_{0}$. 
In the following, we consider each term in the expression 
to deduce one of the four basic laws 
of the one-dimensional two-layer shallow-water equations. 

Let us first discuss the term representing 
the rate of change of the circulation around $\Gamma_{0}$, 
i.e.\ the term on the left-hand side of (\ref{2.200}). 
Using (\ref{3.30}), we have 
\begin{align}
  \label{3.70}
  \frac{\mathrm{d}}{\mathrm{d}t_{\ast}}\oint_{\Gamma_{0}}
       \boldsymbol{u}_{\ast}\cdot\boldsymbol{t}\:\mathrm{d}s_{\ast}
 &=
  \beta gH\frac{\mathrm{d}}{\mathrm{d}t}
    \left\{
       \int_{x_{2}}^{x_{1}}
       \left(\left.u\right|_{z=\mbox{}+0}
            -\left.u\right|_{z=1-0}
       \right)\:\mathrm{d}x
    \right.
    \notag\\
 &\quad
       +\left(\frac{H}{L}\right)^{2}
    \left.
       \int_{\!\mbox{}+0}^{1-0}
       \left(\left.w\right|_{x=x_{1}}
            -\left.w\right|_{x=x_{2}}
       \right)\:\mathrm{d}z
    \right\}.
\end{align}
However, since $H/L\ll 1$, the second term in the braces, 
i.e.\ the contributions from $BC$ and $DA$, should be neglected. 
Hence, if (\ref{3.50}) is used, 
(\ref{3.70}) can be approximated by 
\begin{equation}
  \label{3.80}
  \frac{\mathrm{d}}{\mathrm{d}t_{\ast}}\oint_{\Gamma_{0}}
       \boldsymbol{u}_{\ast}\cdot\boldsymbol{t}\:\mathrm{d}s_{\ast}
 =\beta gH\frac{\mathrm{d}}{\mathrm{d}t}
       \int_{x_{2}}^{x_{1}}(u_{1}-u_{2})\:\mathrm{d}x.
\end{equation}

We next consider the term representing 
the rate of advection of circulation across $\Gamma_{0}$, 
i.e.\ the first term on the right-hand side of (\ref{2.200}). 
Since $\boldsymbol{q}_{\ast}=\boldsymbol{u}_{\ast}$ on $\Gamma_{0}$, we can write 
\begin{align}
  \label{3.90}
  -\oint_{\Gamma_{0}}\omega_{n\ast}
       \boldsymbol{q}_{\ast}\cdot\boldsymbol{\nu}\:\mathrm{d}s_{\ast}=
 &-
   \beta gH
    \left\{
       \int_{x_{2}}^{x_{1}}
       \left(\left.w\frac{\partial u}{\partial z}\right|_{z=\mbox{}+0}
            -\left.w\frac{\partial u}{\partial z}\right|_{z=1-0}
       \right)\:\mathrm{d}x
    \right.
    \notag\\
 &-
    \left.
       \int_{\!\mbox{}+0}^{1-0}
       \left(\left.u\frac{\partial u}{\partial z}\right|_{x=x_{1}}
            -\left.u\frac{\partial u}{\partial z}\right|_{x=x_{2}}
       \right)\:\mathrm{d}z
    \right\},
\end{align}
where we have used (\ref{3.30}) and (\ref{3.40}). 
It should be noted that the first term in the braces 
represents the advection of circulation across $AB$ and $CD$. 
Since we have assumed that the upper and lower vortex sheets 
never separate from the boundaries, 
this advection of circulation is in fact absent. 
Thus, using the identity 
$u\partial u/\partial z=\partial(\frac{1}{2}u^2)/\partial z$, we obtain 
\begin{equation}
  \label{3.100}
  -\oint_{\Gamma_{0}}\omega_{n\ast}
       \boldsymbol{q}_{\ast}\cdot\boldsymbol{\nu}\:\mathrm{d}s_{\ast}
 =-\beta gH
    \left[
       \tfrac{1}{2} u_{1}^{2}-\tfrac{1}{2} u_{2}^{2}
    \right]_{x=x_{2}}^{x=x_{1}}.
\end{equation}
Here we have introduced the notation 
$[\varphi(x,t)]_{x=x_{2}}^{x=x_{1}}
=\varphi(x_{1},t)-\varphi(x_{2},t)$. 

We can also rewrite the term representing the rate of generation 
of circulation due to baroclinicity, i.e.\ the second term 
on the right-hand side of (\ref{2.200}), using (\ref{3.60}) and 
\begin{equation}
  \label{3.110}
  -\oint_{\Gamma_{0}}\frac{1}{\rho_{\ast}}
               \nabla p_{\ast}\cdot\boldsymbol{t}\:\mathrm{d}s_{\ast} 
 =-\left(\frac{1}{\rho_{1}}-\frac{1}{\rho_{2}}\right)
   \left\{
     \left.p_{\ast}\right|_{x=x_{1},z=h(x_{1},t)}
    -\left.p_{\ast}\right|_{x=x_{2},z=h(x_{2},t)}
   \right\}.
\end{equation}
Indeed, substitution of (\ref{3.60}) into (\ref{3.110}) leads to 
\begin{equation}
  \label{3.120}
  -\oint_{\Gamma_{0}}\frac{1}{\rho_{\ast}}
               \nabla p_{\ast}\cdot\boldsymbol{t}\:\mathrm{d}s_{\ast} 
 =-\beta gH\left[h-\beta\eta\right]_{x=x_{2}}^{x=x_{1}}.
\end{equation}

On the other hand, since the external force 
acting on the fluid is the force of gravity, 
the term representing the rate of generation of circulation 
due to the external force, 
i.e.\ the third term on the right-hand side of (\ref{2.200}), 
vanishes identically: 
\begin{equation}
  \label{3.130}
  \oint_{\Gamma_{0}}\boldsymbol{f}_{\ast}\cdot\boldsymbol{t}\:\mathrm{d}s_{\ast}=0.
\end{equation}

Finally, let us consider the term representing 
the rate of diffusion of circulation across $\Gamma_{0}$, 
i.e.\ the last term on the right-hand side of (\ref{2.200}). 
Since this term is the line integral 
along $\Gamma_{0}$ of the viscous force per unit mass, 
it is seen that the main contributions to the term arise from 
the points of intersection of $\Gamma_{0}$ 
and the vortex sheet coincident with the interface. 
Inside this interfacial vortex sheet, 
the viscous force per unit mass is expected to be 
of the same order of magnitude as the inertia force per unit mass: 
\begin{equation}
  \label{3.140}
  \left|\frac{1}{\rho_{\ast}}\boldsymbol{F}_{\ast}\right|
 ={\it O}\left(\frac{\beta gH}{L}\right),
  \qquad |y|<\tfrac{1}{2},\quad z=h.
\end{equation}
Thus, if $\delta$ is the scale characteristic of the dimensional 
thickness of the interfacial vortex sheet, 
the magnitude of the term can be estimated as follows: 
\begin{equation}
  \label{3.150}
  \left|\oint_{\Gamma_{0}}\frac{1}{\rho_{\ast}}\boldsymbol{F}_{\ast}
              \cdot\boldsymbol{t}\:\mathrm{d}s_{\ast}\right|
 \le\oint_{\Gamma_{0}}\left|\frac{1}{\rho_{\ast}}\boldsymbol{F}_{\ast}\right|
              \:\mathrm{d}s_{\ast}
 ={\it O}\left(\beta gH\frac{\delta}{L}\right).
\end{equation}
Since $\delta/L\ll 1$, it follows from (\ref{3.150}) 
that the term representing the rate of diffusion of circulation 
across $\Gamma_{0}$ is negligible in comparison with 
(\ref{3.80}), (\ref{3.100}), and (\ref{3.120}). 

As a result, by applying the conservation law of circulation 
to $\Gamma_{0}$, we obtain 
\begin{equation}
  \label{3.160}
  \frac{\mathrm{d}}{\mathrm{d}t}\int_{x_{2}}^{x_{1}}
          (u_{1}-u_{2})\:\mathrm{d}x
=-\left[
          \tfrac{1}{2} u_{1}^{2}-\tfrac{1}{2} u_{2}^{2}+h-\beta\eta
  \right]_{x=x_{2}}^{x=x_{1}}.
\end{equation}
What is shown by (\ref{3.160}) is that 
the rate of change of the circulation around $\Gamma_{0}$ 
is equal to the sum of 
the rate of advection of circulation across $\Gamma_{0}$ and 
the rate of generation of circulation due to baroclinicity; 
this is one of the four basic laws of 
the one-dimensional two-layer shallow-water equations 
(the `fourth' basic law stated in \S\,1). 

We note here that, in the region enclosed by $\Gamma_{0}$, 
the surface density of circulation $\omega_{n\ast}$ 
takes nonzero values only inside the interfacial vortex sheet. 
Thus it follows that, in the region, circulation is contained 
only inside the interfacial vortex sheet. 
This allows us to interpret the above law as describing 
the balance of circulation for the interfacial vortex sheet. 
For this reason, the above law is henceforth referred to as 
the conservation law of circulation for the interfacial vortex sheet. 

\subsubsection{Conservation law of momentum 
               for the upper and lower layers together}
We next consider the volume $\Omega_{0}$ 
consisting of the upper and lower layers 
between the planes $x=x_{1}$ and $x=x_{2}$ 
but out of the vortex sheets on the boundaries: 
\[
  \Omega_{0}\!:\quad x_{2}\le x\le x_{1},
               \quad |y|\le\tfrac{1}{2}-0,
               \quad \mbox{}+0\le z\le 1-0.
\]
By applying the conservation law of momentum 
in the $x$-direction to $\Omega_{0}$, we have 
\begin{align}
  \label{3.170}
 &
 \frac{\mathrm{d}}{\mathrm{d}t}\int_{x_{2}}^{x_{1}}
 \left\{
          hu_{1}+(1-\beta)(1-h)u_{2}\right\}\:\mathrm{d}x
  \notag\\
&\qquad\quad=
-\left[
   hu_{1}^{2}+(1-\beta)(1-h)u_{2}^{2}+\tfrac{1}{2} h^{2}+(1-\beta)\eta
 \right]_{x=x_{2}}^{x=x_{1}}.
\end{align}
Equation (\ref{3.170}) states that the rate of change 
of the $x$-component of the momentum in $\Omega_{0}$, 
represented by the left-hand side, 
is equal to the sum of 
the rate of advection of the $x$-component of momentum 
across the surface enclosing $\Omega_{0}$ and 
the $x$-component of the total pressure force acting on the surface. 
This is the conservation law of momentum 
for the upper and lower layers together. 

\subsubsection{Conservation laws of mass 
               for the upper layer and for the lower layer}
Finally, we consider two other volumes in the channel. 
One of them, which is denoted by $\Omega_{1}$, 
is the volume of the lower layer defined by 
\[
  \Omega_{1}\!:\quad x_{2}\le x\le x_{1},
               \quad |y|\le\tfrac{1}{2}-0,
               \quad \mbox{}+0\le z\le h-0.
\]
The other volume, which is denoted by $\Omega_{2}$, 
is that of the upper layer defined by 
\[
  \Omega_{2}\!:\quad x_{2}\le x\le x_{1},
               \quad |y|\le\tfrac{1}{2}-0,
               \quad h+0\le z\le 1-0.
\]
By applying the conservation law of mass to $\Omega_{1}$, we obtain 
\begin{equation}
  \label{3.180}
  \frac{\mathrm{d}}{\mathrm{d}t}\int_{x_{2}}^{x_{1}}h\:\mathrm{d}x
=-\left[hu_{1}\right]_{x=x_{2}}^{x=x_{1}},
\end{equation}
i.e.\ the rate of change of the mass in $\Omega_{1}$ 
is equal to the rate of advection of mass 
across the surface enclosing $\Omega_{1}$. 
Equation (\ref{3.180}) expresses 
the conservation law of mass for the lower layer. 
Similarly, the rate of change of the mass in $\Omega_{2}$ 
must be equal to the rate of advection of mass 
across the surface enclosing $\Omega_{2}$: 
\begin{equation}
  \label{3.190}
  \frac{\mathrm{d}}{\mathrm{d}t}\int_{x_{2}}^{x_{1}}
       (1-\beta)(1-h)\:\mathrm{d}x
=-\left[(1-\beta)(1-h)u_{2}\right]_{x=x_{2}}^{x=x_{1}}.
\end{equation}
This equation expresses the conservation law 
of mass for the upper layer. 

\subsection{Derivation of the equations}
We now wish to explain how 
the one-dimensional two-layer shallow-water equations 
are derived from the above four basic laws. 
Suppose first that the variables $u_{1}$, $u_{2}$, 
$h$, and $\eta$ are all continuously differentiable. 
Then the four basic laws can be expressed 
by partial differential equations 
equivalent to (\ref{3.160})--(\ref{3.190}): 
the conservation law of circulation 
for the interfacial vortex sheet can be expressed by 
\begin{equation}
  \label{3.200}
  \frac{\partial}{\partial t}(u_{1}-u_{2})
=-\frac{\partial}{\partial x}
  (
  \tfrac{1}{2} u_{1}^{2}-\tfrac{1}{2} u_{2}^{2}+h-\beta\eta
   );
\end{equation}
the conservation law of momentum 
for the upper and lower layers together by 
\begin{align}
  \label{3.210}
  &
  \frac{\partial}{\partial t}\left\{hu_{1}+(1-\beta)(1-h)u_{2}\right\}
  \notag\\
 &\qquad=
  -\frac{\partial}{\partial x}
   \left\{
        hu_{1}^{2}+(1-\beta)(1-h)u_{2}^{2}+\tfrac{1}{2} h^{2}+(1-\beta)\eta
   \right\};
\end{align}
the conservation law of mass for the lower layer by 
\begin{equation}
  \label{3.220}
  \frac{\partial h}{\partial t}
=-\frac{\partial}{\partial x}(hu_{1});
\end{equation}
and the conservation law of mass for the upper layer by 
\begin{equation}
  \label{3.230}
  \frac{\partial}{\partial t}\left\{(1-\beta)(1-h)\right\}
=-\frac{\partial}{\partial x}\left\{(1-\beta)(1-h)u_{2}\right\}.
\end{equation}
After some manipulations, we can show 
that (\ref{3.200})--(\ref{3.230}) yield 
\begin{equation}
  \label{3.240}
  \left.
  \begin{array}{c}
  \displaystyle
  \frac{\partial u_{1}}{\partial t}+u_{1}\frac{\partial u_{1}}{\partial x}
=-\frac{\partial h}{\partial x}-(1-\beta)\frac{\partial \eta}{\partial x},
  \quad
  \frac{\partial u_{2}}{\partial t}+u_{2}\frac{\partial u_{2}}{\partial x}
=-\frac{\partial \eta}{\partial x},\\[10pt]
  \displaystyle
  \frac{\partial h}{\partial t}
=-\frac{\partial}{\partial x}(hu_{1}),
  \quad
  \frac{\partial}{\partial t}(1-h)
=-\frac{\partial}{\partial x}\left\{(1-h)u_{2}\right\}.
  \end{array}
  \right\}
\end{equation}
This is a well-known form of
the one-dimensional two-layer shallow-water equations. 

The special case in which the density in the upper layer 
is much smaller than that in the lower layer, 
i.e.\ the case in which $\beta=1-0$, 
deserves to be considered separately. 
In this case, the expression (\ref{3.210}) 
of the conservation law of momentum for 
the upper and lower layers together is approximated by 
\begin{equation}
  \label{3.250}
  \frac{\partial}{\partial t}(hu_{1})
=-\frac{\partial}{\partial x}(hu_{1}^{2}+\tfrac{1}{2} h^{2}).
\end{equation}
As a consequence, from (\ref{3.250}) and the expression (\ref{3.220}) 
of the conservation law of mass for the lower layer, 
we find the following system of equations for $u_{1}$ and $h$: 
\begin{equation}
  \label{3.260}
  \frac{\partial u_{1}}{\partial t}+u_{1}\frac{\partial u_{1}}{\partial x}
=-\frac{\partial h}{\partial x},
  \quad
  \frac{\partial h}{\partial t}
=-\frac{\partial}{\partial x}(hu_{1}).
\end{equation}
This is obviously a form of 
the one-dimensional single-layer shallow-water equations. 

The same system of equations as (\ref{3.260}) arises 
in a different context. 
To show this, we first assume that 
the thickness of the lower layer is sufficiently small 
compared with the depth of the channel, i.e.\ $h\ll 1$. 
When $h\ll 1$, the expression (\ref{3.230}) of the conservation law 
of mass for the upper layer reduces to $\partial u_{2}/\partial x=0$. 
This allows us to put 
\begin{equation}
  \label{3.270}
  u_{2}=\mbox{constant}.
\end{equation}
If, in addition, it is assumed that the density difference 
between the layers is very small compared with 
the density in the lower layer, i.e.\ $\beta=\!\mbox{}+0$, 
then the expression (\ref{3.200}) of the conservation law of circulation 
for the interfacial vortex sheet becomes 
\begin{equation}
  \label{3.280}
  \frac{\partial u_{1}}{\partial t}
=-\frac{\partial}{\partial x}
  (
  \tfrac{1}{2} u_{1}^{2}+h
   ).
\end{equation}
Coupling (\ref{3.280}) with the expression (\ref{3.220}) 
of the conservation law of mass for the lower layer, 
we obtain the following system of equations: 
\begin{equation}
  \label{3.290}
  \frac{\partial u_{1}}{\partial t}+u_{1}\frac{\partial u_{1}}{\partial x}
=-\frac{\partial h}{\partial x},
  \quad
  \frac{\partial h}{\partial t}
=-\frac{\partial}{\partial x}(hu_{1}).
\end{equation}
This is exactly the same system of equations as (\ref{3.260}). 

However, we need to recognize that 
(\ref{3.290}) represents, 
under the condition (\ref{3.270}), 
the conservation laws of mass for the lower layer 
and of circulation for the interfacial vortex sheet. 
On the other hand, (\ref{3.260}) represents 
the conservation laws of mass for the lower layer 
and of momentum for the upper and lower layers together. 
From this instance, we can understand 
that mathematically equivalent systems 
of partial differential equations do not necessarily 
represent the same basic physical laws. 
Though this fact seems to be almost obvious, 
it is liable to be overlooked.

%% file: section4.tex
\section{Internal bores}
As stated in \S\,1, within the framework of 
the one-dimensional two-layer shallow-water equations, 
an internal bore in a two-fluid system 
is represented by a shock; 
the primary objective of this section 
is to formulate the shock conditions for such a shock. 
They are shown to be derived from the four basic laws 
of the equations elucidated in \S\,3. 
It must be emphasized, however, 
that not every shock satisfying the shock conditions 
represents a bore that can actually exist; 
if a shock represents a real bore, 
then the shock satisfies two additional conditions. 
These conditions, which we call the energy condition 
and the evolutionary condition, are also discussed. 
Finally, it is shown that the shock conditions 
can satisfactorily account for the results 
of the experiments by Wood \& Simpson (1984) 
on internal bores advancing into 
two stationary layers of fluid. 

\subsection{Shock conditions}
We consider the same physical situation as that in \S\,3. 
However, we now suppose 
that an internal bore moving parallel to 
the side boundaries is present in the channel. 
Except the inside of the bore, 
the motion of the fluid is assumed to possess 
the properties described in \S\,3; 
the motion changes rapidly inside the bore 
from the state on one side of the bore 
to that on the other side. 
In order to represent properly the motion outside the bore, 
we use again the dimensionless coordinates $x$, $y$, $z$ 
and the dimensionless time $t$. 

Let $l_{s}$ be the scale 
of the dimensional distance measured across the bore. 
We assume that $l_{s}$ is very small compared with 
the length scale $L$ of the motion 
outside the bore, i.e.\ $l_{s}/L\ll 1$. 
Then the bore may be identified, 
in the dimensionless coordinate system, 
with a plane normal to the $x$-axis 
across which the motion changes discontinuously. 
We denote the $x$-coordinate of this plane by $x_{s}(t)$. 

For $x\neq x_{s}$, 
the motion of the fluid is specified in terms of 
the variables $u_{1}$, $u_{2}$, $h$, and $\eta$ 
introduced in \S\,3. 
These variables are discontinuous at $x=x_{s}$, 
for the motion changes discontinuously there. 
On the other hand, as we see from the argument in \S\,3, 
the set of the variables is determined 
in each of the infinite intervals 
$x\le x_{s}-0$ and $x\ge x_{s}+0$ 
as a continuous solution of 
the one-dimensional two-layer shallow-water equations. 
Thus it follows that, within the framework of the equations, 
the bore is represented by a shock, 
i.e.\ a discontinuity dividing two continuous solutions, 
located at $x=x_{s}$. 

When the bore is represented in this way, 
the variables $u_{1}$, $u_{2}$, $h$, and $\eta$ 
at $x=x_{s}-0$ and at $x=x_{s}+0$ 
need to be related by some conditions. 
Our aim is to formulate these shock conditions. 
As explained in the following, 
each of them is derived 
from one of the four basic laws 
of the one-dimensional two-layer shallow-water equations. 

First, let us focus our attention on the plane $y=0$, 
and consider the closed curve $\Gamma_{0}$ introduced in \S\,3. 
We suppose that the arbitrary constants $x_{1}$ and $x_{2}$ 
are so chosen that $x_{1}>x_{s}>x_{2}$. 
Thus the bore lies between $x=x_{1}$ and $x=x_{2}$, 
as shown in figure 2. 
\begin{figure}[tb]
 \centering
 \includegraphics{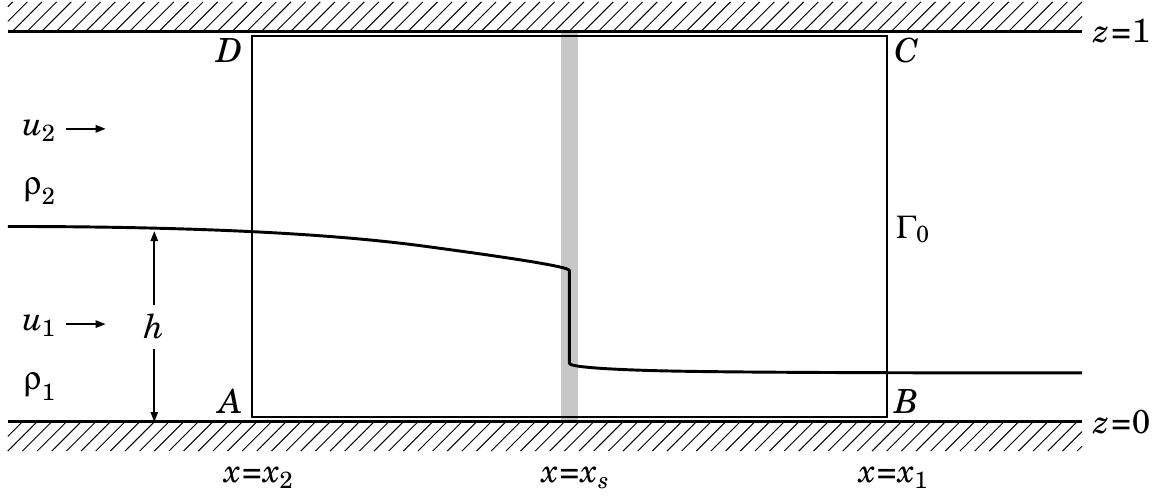}
 \caption[]{Position of the internal bore relative to $\Gamma_{0}$.
            The bore is identified with a plane across which
            the fluid motion changes discontinuously;
            the grey band shows this plane.}
\end{figure}

However, we assume that the conservation law of circulation 
for the interfacial vortex sheet is not altered 
by the presence of the bore. 
Then it follows that, 
while the bore lies between $x=x_{1}$ and $x=x_{2}$, 
the rate of change of the circulation around $\Gamma_{0}$ 
is still equal to the sum of 
the rate of advection of circulation across $\Gamma_{0}$ and 
the rate of generation of circulation due to baroclinicity. 
We can formulate a shock condition on the basis of this law. 
To this end, we must first express the law mathematically. 
It is expressed by (\ref{3.160}) if the bore is absent; 
we wish to examine how this expression should be modified. 

The rate of change of the circulation around $\Gamma_{0}$ 
is now expressed by 
\begin{align}
  \label{4.10}
  \frac{\mathrm{d}}{\mathrm{d}t_{\ast}}\oint_{\Gamma_{0}}
       \boldsymbol{u}_{\ast}\cdot\boldsymbol{t}\:\mathrm{d}s_{\ast}
 &=
    \beta gH\frac{\mathrm{d}}{\mathrm{d}t}
    \left\{
       \int_{x_{2}}^{x_{s}-0}(u_{1}-u_{2})\:\mathrm{d}x
      +\int_{x_{s}+0}^{x_{1}}(u_{1}-u_{2})\:\mathrm{d}x
    \right\}
    \notag\\
 &\quad
   +\beta gH\frac{\mathrm{d}}{\mathrm{d}t}
       \int_{x_{s}-0}^{x_{s}+0}
       \frac{\left.\boldsymbol{u}_{\ast}\right|_{z=\mbox{}+0}
            -\left.\boldsymbol{u}_{\ast}\right|_{z=1-0}}
            {(\beta gH)^{\frac{1}{2}}}
       \cdot\boldsymbol{i}\:\mathrm{d}x
\end{align}
in place of (\ref{3.80}). 
The last term represents 
the contribution from the inside of the bore. 
However, we may expect the velocity there to have 
the same order of magnitude as that outside the bore: 
$|\boldsymbol{u}_{\ast}|/(\beta gH)^{\frac{1}{2}}={\it O}(1)$ 
for $x_{s}-0\le x\le x_{s}+0$. 
Hence the magnitude of the integrand 
in the term may be estimated to be ${\it O}(1)$. 
Since the integral is taken over an infinitesimal interval, 
this estimate reveals that the term is negligible. 
Accordingly, if the integral on the right-hand side of (\ref{3.80}) 
is interpreted as the improper integral 
\begin{equation}
  \label{4.20}
  \int_{x_{2}}^{x_{1}}(u_{1}-u_{2})\:\mathrm{d}x
 =\int_{x_{2}}^{x_{s}-0}(u_{1}-u_{2})\:\mathrm{d}x
 +\int_{x_{s}+0}^{x_{1}}(u_{1}-u_{2})\:\mathrm{d}x,
\end{equation}
the rate of change of the circulation around $\Gamma_{0}$ 
is again expressed by (\ref{3.80}). 

On the other hand, the rate of advection 
of circulation across $\Gamma_{0}$ 
is now given by 
\begin{align}
  \label{4.30}
    -\oint_{\Gamma_{0}}\omega_{n\ast}
        \boldsymbol{q}_{\ast}\cdot\boldsymbol{\nu}\:\mathrm{d}s_{\ast}=
 &-
     \beta gH
     \left[
        \tfrac{1}{2} u_{1}^{2}-\tfrac{1}{2} u_{2}^{2}
     \right]_{x=x_{2}}^{x=x_{1}}
     \notag\\
 &-
     \beta gH
     \int_{x_{s}-0}^{x_{s}+0}
     \frac{\left.\omega_{n\ast}\boldsymbol{u}_{\ast}\right|_{z=1-0}
          -\left.\omega_{n\ast}\boldsymbol{u}_{\ast}\right|_{z=\mbox{}+0}}
          {\beta gH/L}
     \cdot\boldsymbol{k}\:\mathrm{d}x
\end{align}
in place of (\ref{3.100}). 
The last term in this expression has been introduced 
to represent the advection of circulation 
across $\Gamma_{0}$ that occurs inside the bore. 
However, since the vortex sheets on the upper and lower boundaries 
have been assumed never to separate from the boundaries, 
this term vanishes identically. 
Thus, so long as this assumption is valid, 
the rate of advection of circulation 
across $\Gamma_{0}$ is again expressed by (\ref{3.100}). 

Also, it can be shown that the rate of generation of circulation 
due to baroclinicity is still expressed by (\ref{3.120}). 
Thus we reach the conclusion that, 
while the bore lies between $x=x_{1}$ and $x=x_{2}$, 
the conservation law of circulation for 
the interfacial vortex sheet is expressed by (\ref{3.160}) 
without any modifications in the form; 
however, the integral in this expression must be interpreted 
as the improper integral defined by (\ref{4.20}). 

From this expression of the law, 
we can now formulate the desired shock condition. 
If the procedure described by Whitham (1974, \S\,5.8)
is used on (\ref{3.160}), it is found that the shock 
representing the bore satisfies the condition 
\begin{equation}
  \label{4.40}
  -U_{s}\left[u_{1}-u_{2}\right]
  =-\left[\tfrac{1}{2} u_{1}^{2}-\tfrac{1}{2} u_{2}^{2}+h-\beta\eta\right].
\end{equation}
Here $U_{s}(t)=\mathrm{d}x_{s}/\mathrm{d}t$ 
is the velocity of the shock, and 
$[\varphi(x,t)]=\varphi(x_{s}+0,t)-\varphi(x_{s}-0,t)$ 
denotes the jump in $\varphi(x,t)$ across the shock. 
This is the shock condition corresponding to 
the conservation law of circulation 
for the interfacial vortex sheet. 

Similarly, the remaining shock conditions can be obtained 
from the other three basic laws of the equations: 
from the conservation law of momentum 
for the upper and lower layers together, we obtain 
\begin{equation}
  \label{4.50}
  -U_{s}\left[hu_{1}+(1-\beta)(1-h)u_{2}\right]
  =-\left[
     hu_{1}^{2}+(1-\beta)(1-h)u_{2}^{2}+\tfrac{1}{2} h^{2}+(1-\beta)\eta
    \right];
\end{equation}
from the conservation law of mass for the lower layer, 
\begin{equation}
  \label{4.60}
  -U_{s}\left[h\right]=-\left[hu_{1}\right];
\end{equation}
and from the conservation law of mass for the upper layer, 
\begin{equation}
  \label{4.70}
  -U_{s}\left[(1-\beta)(1-h)\right]=-\left[(1-\beta)(1-h)u_{2}\right].
\end{equation}
The conditions (\ref{4.40})--(\ref{4.70}) 
constitute the set of shock conditions for 
the one-dimensional two-layer shallow-water equations (\ref{3.240}). 

It is also possible to write the shock conditions 
in terms of the relative velocities 
\begin{equation}
  \label{4.80}
  v_{1}=u_{1}-U_{s},\quad v_{2}=u_{2}-U_{s}.
\end{equation}
Substituting (\ref{4.80}) into (\ref{4.40})--(\ref{4.70}) 
and rearranging the results, we have 
\begin{equation}
  \label{4.90}
  \left.
  \begin{array}{c}
  \displaystyle
  \left[\tfrac{1}{2} v_{1}^{2}-\tfrac{1}{2} v_{2}^{2}+h-\beta\eta\right]=0,
  \quad
  \left[hv_{1}\right]=0,
  \quad
  \left[(1-\beta)(1-h)v_{2}\right]=0,\\[6pt]
  \displaystyle
  \left[
    hv_{1}^{2}+(1-\beta)(1-h)v_{2}^{2}+\tfrac{1}{2} h^{2}+(1-\beta)\eta
  \right]=0.
  \end{array}
  \right\}
\end{equation}

Let us confirm that the shock conditions (\ref{4.40})--(\ref{4.70}) 
are consistent with the theory of bores in classical hydraulics. 
To this end, we now suppose that the density in the upper layer 
is much smaller than that in the lower layer, i.e.\ $\beta=1-0$. 
Then the following set of shock conditions 
on $u_{1}$ and $h$ can be found: 
\begin{equation}
  \label{4.100}
  -U_{s}\left[hu_{1}\right]=-\left[hu_{1}^{2}+\tfrac{1}{2} h^{2}\right],
  \quad
  -U_{s}\left[h\right]=-\left[hu_{1}\right].
\end{equation}
The former condition of (\ref{4.100}) is obtained 
from the shock condition (\ref{4.50}) 
corresponding to the conservation law of momentum 
for the upper and lower layers together; 
the latter is the shock condition (\ref{4.60}) 
corresponding to the conservation law of mass for the lower layer. 
The set of shock conditions (\ref{4.100}) 
is the one well known in the theory of bores 
in classical hydraulics (see Whitham 1974, \S\,13.10). 

Note that, when $\beta=1-0$, the variables $u_{1}$ and $h$ 
are governed in each of the infinite intervals 
$x\le x_{s}-0$ and $x\ge x_{s}+0$ by the one-dimensional 
single-layer shallow-water equations (\ref{3.260}). 
This implies that (\ref{4.100}) is 
the set of shock conditions for (\ref{3.260}). 
Next, let us direct our attention to 
the system of equations (\ref{3.290}). 
This is mathematically 
the same system of equations as (\ref{3.260}). 
However, we can see from the discussion below that (\ref{3.290}) 
requires a set of shock conditions different from (\ref{4.100}). 

When deriving (\ref{3.290}), we assumed that 
the thickness of the lower layer is sufficiently small 
compared with the depth of the channel, i.e.\ $h\ll 1$. 
If $h\ll 1$, the shock condition (\ref{4.70}) 
corresponding to the conservation law of mass 
for the upper layer reduces to 
\begin{equation}
  \label{4.110}
  \left[u_{2}\right]=0.
\end{equation}
Furthermore, it was assumed that the density difference 
between the layers is very small compared with 
the density in the lower layer, i.e.\ $\beta=\!\mbox{}+0$. 
Let us put again $\beta=\!\mbox{}+0$ and 
substitute (\ref{4.110}) into the shock condition (\ref{4.40}) 
corresponding to the conservation law of circulation 
for the interfacial vortex sheet; 
coupling the result with the shock condition (\ref{4.60}) corresponding 
to the conservation law of mass for the lower layer, we obtain 
\begin{equation}
  \label{4.120}
  -U_{s}\left[u_{1}\right]=-\left[\tfrac{1}{2} u_{1}^{2}+h\right],
  \quad
  -U_{s}\left[h\right]=-\left[hu_{1}\right].
\end{equation}
This is the set of shock conditions for (\ref{3.290}), 
and is obviously different from (\ref{4.100}). 

This result may seem somewhat paradoxical 
since there is no mathematical difference between 
(\ref{3.260}) and (\ref{3.290}). 
We should realize, however, that each of the shock conditions 
for a system of equations is derived from 
one of the basic physical laws that the system represents. 
As was pointed out at the end of \S\,3.2, 
mathematically equivalent systems of equations 
do not necessarily represent the same basic physical laws. 
Hence, even though two systems of equations 
are mathematically equivalent, 
the sets of shock conditions for the systems may be different. 
The above result is an instance of this fact. 

\subsection{Energy condition}
It has already been shown that, within the framework of 
the one-dimensional two-layer shallow-water equations, 
an internal bore is represented by a shock 
satisfying the shock conditions (\ref{4.40})--(\ref{4.70}). 
We should note, however, that there are shocks which 
satisfy the shock conditions but do not correspond to real bores. 
In order to exclude such spurious shocks, 
we need to impose two additional conditions. 
Our next aim is to elucidate these additional conditions. 
We first formulate the condition derived from the constraint 
that mechanical energy cannot be generated inside a bore. 

We consider the internal bore in \S\,4.1, 
and introduce the following quantity 
expressed in terms of the relative velocities (\ref{4.80}): 
\begin{equation}
  \label{4.150}
   \Phi_{s}(t)=
 -\left[hv_{1}\left\{\tfrac{1}{2} v_{1}^{2}+h+(1-\beta)\eta\right\}\right]
 -\left[(1-\beta)(1-h)v_{2}(\tfrac{1}{2} v_{2}^{2}+\eta)\right].
\end{equation}
It can readily be shown 
that $\rho_{1}(\beta gH)^{\frac{3}{2}}HW\Phi_{s}$ 
gives the rate of dissipation of mechanical energy inside the bore. 
Since mechanical energy cannot be generated inside the bore, 
we see that the shock representing the bore 
satisfies the condition 
\begin{equation}
  \label{4.140}
     \Phi_{s}\ge 0.
\end{equation}
We call this condition the energy condition. 
It is to be regarded as a necessary condition 
for a shock satisfying the shock conditions 
to represent a real bore. 

Now, referring to (\ref{4.90}), we observe that 
\begin{equation}
  \label{4.160}
  \left.\left\{hv_{1}+(1-\beta)(1-h)v_{2}\right\}\right|_{x=x_{s}-0}
 =\left.\left\{hv_{1}+(1-\beta)(1-h)v_{2}\right\}\right|_{x=x_{s}+0}
 =I_{s}(t).
\end{equation}
Here $I_{s}$ represents 
the rate of total advection of mass across the bore. 
In terms of $I_{s}$, $\Phi_{s}$ can be rewritten, 
by the use of (\ref{4.90}), as follows: 
\begin{equation}
  \label{4.170}
  \Phi_{s}=-I_{s}\left[\tfrac{1}{2} v_{1}^{2}+h+(1-\beta)\eta\right]
          =-I_{s}\left[\tfrac{1}{2} v_{2}^{2}+\eta\right].
\end{equation}
This can be seen if we note 
that the first shock condition in (\ref{4.90}), 
which corresponds to the conservation law of circulation 
for the interfacial vortex sheet, is equivalent to 
\begin{equation}
  \label{4.190}
  \left[\tfrac{1}{2} v_{1}^{2}+h+(1-\beta)\eta\right]
 =\left[\tfrac{1}{2} v_{2}^{2}+\eta\right].
\end{equation}
This expression states that, 
if viewed by an observer moving with the bore, 
the changes in total head in the two layers are equal. 

In connection with the energy condition, 
we wish to discuss in the following 
the rate of dissipation of mechanical energy 
in each of the two layers. 

To this end, we should realize 
the following fact (see Klemp et al.\ 1997): 
the transfer of mechanical energy between the layers 
may occur inside the bore owing to turbulence. 
We denote by $\rho_{2}(\beta gH)^{\frac{3}{2}}HW\Psi_{s}(t)$ 
the rate of this mechanical energy transfer 
(from the upper layer to the lower layer) 
perceived by an observer moving with the bore. 

Now consider the following quantity: 
\begin{equation}
  \label{4.191}
  \Phi_{s1}(t)=
  -\left[hv_{1}\left\{\tfrac{1}{2} v_{1}^{2}+h+(1-\beta)\eta\right\}\right]
  +(1-\beta)\Psi_{s}.
\end{equation}
We can see that $\rho_{1}(\beta gH)^{\frac{3}{2}}HW\Phi_{s1}$ 
gives the rate of dissipation 
of mechanical energy in the lower layer. 
The corresponding quantity for the upper layer is 
\begin{equation}
  \label{4.192}
  \Phi_{s2}(t)=
  -\left[(1-\beta)(1-h)v_{2}(\tfrac{1}{2} v_{2}^{2}+\eta)\right]
  -(1-\beta)\Psi_{s}.
\end{equation}
As is evident from (\ref{4.150}), 
the sum of these quantities is equal to $\Phi_{s}$. 

It is important to note, however, that 
there is no means of predicting the value of $\Psi_{s}$. 
Hence we cannot predict 
the values of $\Phi_{s1}$ and $\Phi_{s2}$ either. 
This implies that, within the framework of 
the one-dimensional two-layer shallow-water equations, 
the distribution of mechanical energy dissipation 
inside the bore cannot be predicted in general. 
(The only exceptional case is the one in which 
the density in the upper layer 
is much smaller than that in the lower layer, 
i.e.\ the case in which $\beta=1-0$. 
In this case, we can expect that 
the mechanical energy dissipation occurs 
predominantly in the lower layer: 
on the basis of this expectation, 
the scale factor $\rho_{2}(\beta gH)^{\frac{3}{2}}HW$ 
of $\Psi_{s}$ has been chosen.) 

\subsection{Evolutionary condition}
Before starting the discussion of the other additional condition, 
we need to explain some properties of the one-dimensional 
two-layer shallow-water equations (\ref{3.240}). 

Note first that, taking linear combinations of (\ref{3.240}), 
we can find the following system of equations 
for $u_{1}$, $u_{2}$, and $h$: 
\begin{equation}
  \label{4.200}
  \left.
  \begin{array}{c}
  \displaystyle
  \frac{\partial}{\partial x}\left\{hu_{1}+(1-h)u_{2}\right\}=0,
  \quad
  \frac{\partial h}{\partial t}
=-\frac{\partial}{\partial x}(hu_{1}),\\[9pt]
  \displaystyle
  \frac{\partial u_{1}}{\partial t}+u_{1}\frac{\partial u_{1}}{\partial x}
  -(1-\beta)
  \left(
     \frac{\partial u_{2}}{\partial t}+u_{2}\frac{\partial u_{2}}{\partial x}
  \right)
=-\frac{\partial h}{\partial x}.
  \end{array}
  \right\}
\end{equation}
When this system of equations is hyperbolic, 
it has three families of characteristics 
(see e.g.\ Whitham 1974, \S\,5.1). 
One of them is immediately found from 
the first equation in (\ref{4.200}). 
Each characteristic that belongs to this family carries 
information on the rate of total advection of volume 
with an infinite characteristic velocity. 
We denote this family by $C_{vol}$. 
The characteristic velocities 
of the other two families are finite. 
These families correspond to internal waves. 
We denote by $C_{+}$ the family 
with the larger characteristic velocity $\lambda_{+}$, 
and by $C_{-}$ the one 
with the smaller characteristic velocity $\lambda_{-}$. 

Once (\ref{4.200}) is solved, $\eta$ can be determined 
from the second equation in (\ref{3.240}): 
\begin{equation}
  \label{4.210}
    \frac{\partial \eta}{\partial x}
   =-\frac{\partial u_{2}}{\partial t}-u_{2}\frac{\partial u_{2}}{\partial x}.
\end{equation}
When (\ref{4.210}) is regarded as an equation for $\eta$ alone, 
it has one family of characteristics. 
Each characteristic that belongs to this family 
carries information on $\eta$ 
with an infinite characteristic velocity. 
This family of characteristics is denoted by $C_{\eta}$. 

Having finished the preparation, 
we proceed to the discussion of the other additional condition 
necessary for excluding spurious shocks. 

Consider a shock satisfying 
the shock conditions (\ref{4.40})--(\ref{4.70}). 
If this shock represents a real bore, 
then it is evolutionary (see e.g.\ Landau \& Lifshitz 1987, \S\,88). 
That the shock is evolutionary 
means that the following condition is satisfied: 
there are, at any instant, $N-1$ characteristics 
leaving the shock and $M-N$ characteristics 
reaching the shock or moving with the shock. 
Here $N$ denotes the number of the shock conditions, 
and $M$ the number of the variables in the shock conditions. 
We call this condition the evolutionary condition. 
It also is a necessary condition for a shock satisfying 
the shock conditions to represent a real bore, 
and a shock that violates it cannot persist as a single shock. 

Let us now examine the evolutionary condition in more detail. 
First of all, we need to count the number $N$ of the shock conditions 
and the number $M$ of the variables in the shock conditions. 
It is apparent that $N$ is four. 
On the other hand, $M$ is nine: 
$u_{1}$, $u_{2}$, $h$, and $\eta$ 
at $x=x_{s}-0$ and at $x=x_{s}+0$, and $U_{s}$. 
Hence the evolutionary condition can be restated as follows: 
there are, at any instant, three characteristics 
leaving the shock and five characteristics 
reaching the shock or moving with the shock. 

It is important to note, however, 
that the characteristic velocities of 
$C_{vol}$ and $C_{\eta}$ are infinite. 
Accordingly, whether a shock is evolutionary or not, 
two characteristics leaving the shock 
and two characteristics reaching the shock exist at any instant. 

Thus we see that a shock satisfying 
the shock conditions is evolutionary if and only if 
either of the following conditions is satisfied at any instant: 
\begin{equation}
  \label{4.220}
  \left.
  \begin{array}{l}
  \displaystyle
  \lambda_{-}-U_{s}<0\le\lambda_{+}-U_{s}
  \quad \mbox{at} \quad x=x_{s}-0,\\[2pt]
  \displaystyle
  \lambda_{-}-U_{s}\le\lambda_{+}-U_{s}\le 0
  \quad \mbox{at} \quad x=x_{s}+0;
  \end{array}
  \right\}
\end{equation}
or 
\begin{equation}
  \label{4.230}
  \left.
  \begin{array}{l}
  \displaystyle
  0\le\lambda_{-}-U_{s}\le\lambda_{+}-U_{s}
  \quad \mbox{at} \quad x=x_{s}-0,\\[2pt]
  \displaystyle
  \lambda_{-}-U_{s}\le 0<\lambda_{+}-U_{s}
  \quad \mbox{at} \quad x=x_{s}+0.
  \end{array}
  \right\}
\end{equation}
Here $\lambda_{+}-U_{s}$ and $\lambda_{-}-U_{s}$ are 
the characteristic velocities of $C_{+}$ and $C_{-}$ 
relative to the shock; 
they can be calculated from the formulae (see the Appendix) 
\begin{equation}
  \label{4.240}
  \lambda_{\pm}-U_{s}
 =\overline{v}\pm 
  \left\{
  \overline{v}^{2}
 -\frac{(1-h)v_{1}^{2}+(1-\beta)hv_{2}^{2}-h(1-h)}{(1-\beta)h+(1-h)}
  \right\}^{\frac{1}{2}},
\end{equation}
where $v_{1}$ and $v_{2}$ are the relative velocities (\ref{4.80}), 
and $\overline{v}$ is defined by 
\begin{equation}
  \label{4.250}
  \overline{v}=\frac{(1-h)v_{1}+(1-\beta)hv_{2}}{(1-\beta)h+(1-h)}.
\end{equation}

\subsection{Internal bores advancing 
            into two stationary layers of fluid} 
We have now completed the formulation of the shock conditions 
for the one-dimensional two-layer shallow-water equations 
and the additional conditions 
necessary for excluding spurious shocks. 
Our remaining task in this section is to confirm 
that the behaviour of internal bores can really be predicted 
from the shock conditions (\ref{4.40})--(\ref{4.70}). 
This task is done especially about bores advancing into 
two stationary layers of fluid. 

Let us consider a shock representing such a bore. 
In particular, we concentrate on the case in which 
the density difference between the layers 
is very small in comparison with the density 
in the lower layer, i.e.\ the case in which $\beta=\!\mbox{}+0$.  
Then the shock conditions (\ref{4.90}) expressed in terms of 
the relative velocities (\ref{4.80}) can be simplified to 
\begin{equation}
  \label{4.260}
  \left.
  \begin{array}{c}
  \displaystyle
  \left[\tfrac{1}{2} v_{1}^{2}-\tfrac{1}{2} v_{2}^{2}+h\right]=0,
  \quad
  \left[hv_{1}\right]=0,
  \quad
  \left[(1-h)v_{2}\right]=0,\\[6pt]
  \displaystyle
  \left[
    hv_{1}^{2}+(1-h)v_{2}^{2}+\tfrac{1}{2} h^{2}+\eta
  \right]=0.
  \end{array}
  \right\}
\end{equation}
From (\ref{4.260}), we derive in the following 
a formula that gives the velocity of the shock 
as a function of the thicknesses of the lower layer 
ahead of and behind the shock. 

We first assume, without loss of generality, 
that the shock is advancing in the positive $x$-direction. 
Since the layers are stationary ahead of the shock, we then have 
\begin{equation}
  \label{4.270}
  \left.v_{1}\right|_{x=x_{s}+0}
 =\left.v_{2}\right|_{x=x_{s}+0}=-U_{s}.
\end{equation}
Next, we introduce the following notation: 
\begin{equation}
  \label{4.280}
  h_{a}=\left.h\right|_{x=x_{s}+0},
  \quad
  h_{b}=\left.h\right|_{x=x_{s}-0},
\end{equation}
i.e.\ the thicknesses of the lower layer 
ahead of and behind the shock are denoted 
by $h_{a}$ and $h_{b}$ respectively. 
Substituting (\ref{4.270}) and (\ref{4.280}) into 
the first three shock conditions in (\ref{4.260}) and 
solving for $U_{s}$ the resulting system of algebraic equations, 
we obtain 
\begin{equation}
  \label{4.290}
  U_{s}=\left\{
        \frac{2h_{b}^2(1-h_{b})^2}{h_{a}(1-h_{b})+h_{b}(1-h_{a})}
        \right\}^{\frac{1}{2}}.
\end{equation}

Note that (\ref{4.290}) has been derived 
from the first three shock conditions in (\ref{4.260}). 
This implies that the formula is based on 
the conservation laws of mass and of circulation; 
it is independent of the conservation law of momentum. 

The energy condition imposes a restriction on the values 
that $h_{a}$ and $h_{b}$ in (\ref{4.290}) can take. 
To find the restriction, we note that $I_{s}$ 
defined by (\ref{4.160}) is now given by $-U_{s}<0$. 
It follows from this fact and (\ref{4.170}) that 
the energy condition (\ref{4.140}) is equivalent to 
\begin{equation}
  \label{4.300}
  \left[\tfrac{1}{2} v_{1}^{2}+h+\eta\right]\ge 0.
\end{equation}
This inequality can be expressed, 
by the use of (\ref{4.260})--(\ref{4.290}), as follows: 
\begin{equation}
  \label{4.310}
  \frac{-(h_{b}-\tfrac{1}{2})(h_{b}-h_{a})^3}
       {h_{a}(1-h_{b})+h_{b}(1-h_{a})}\ge 0.
\end{equation}
Thus we see that the energy condition imposes 
the following restriction on $h_{a}$ and $h_{b}$: 
\begin{equation}
  \label{4.320}
  h_{a}\le h_{b}\le \tfrac{1}{2}
  \quad \mbox{or} \quad 
  \tfrac{1}{2}\le h_{b}\le h_{a}.
\end{equation}

The values of $h_{a}$ and $h_{b}$ 
are restricted also by the evolutionary condition. 
To find the restriction imposed by this condition, 
we first consider $\overline{v}$ defined by (\ref{4.250}). 
It can easily be verified that $\overline{v}$ takes 
negative values both at $x=x_{s}-0$ and at $x=x_{s}+0$: 
\begin{equation}
  \label{4.330}
  \left.\overline{v}\right|_{x=x_{s}-0}<0,
  \quad
  \left.\overline{v}\right|_{x=x_{s}+0}<0.
\end{equation}
As we have already seen, the evolutionary condition requires 
that either (\ref{4.220}) or (\ref{4.230}) 
should hold at any instant. 
However, on account of (\ref{4.330}), 
(\ref{4.230}) can never be satisfied. 
Thus, for the shock to be evolutionary, 
(\ref{4.220}) needs to be satisfied. 
Using (\ref{4.260})--(\ref{4.290}), we can show 
that (\ref{4.220}) yields the following restriction 
on $h_{a}$ and $h_{b}$: 
\begin{equation}
  \label{4.340}
  \left.
  \begin{array}{c}
  \displaystyle
  (h_{b}-h_{a})
  \left\{
  (6h_{b}^2-6h_{b}+2)h_{a}-(3h_{b}^2-h_{b})
  \right\}\ge 0,\\[6pt]
  \displaystyle
  (h_{b}-h_{a})
  \left\{
  2h_{b}^3+(2h_{a}-4)h_{b}^2+(2h_{a}^2-4h_{a}+2)h_{b}-(h_{a}^2-h_{a})
  \right\}\ge 0.
  \end{array}
  \right\}
\end{equation}

We note here that the former part of (\ref{4.340}) 
is equivalent to the following restriction: 
\begin{equation}
  \label{4.350}
  \frac{3h_{b}^2-h_{b}}{6h_{b}^2-6h_{b}+2}\le h_{a}\le h_{b}
  \quad \mbox{or} \quad 
  h_{b}\le h_{a}\le \frac{3h_{b}^2-h_{b}}{6h_{b}^2-6h_{b}+2}.
\end{equation}
From this restriction, (\ref{4.320}) follows immediately. 
It can readily be demonstrated, on the other hand, 
that the latter part of (\ref{4.340}) 
is automatically satisfied under (\ref{4.320}). 
We can therefore conclude that (\ref{4.350}) is 
the most stringent restriction on $h_{a}$ and $h_{b}$. 

Now, with this restriction on $h_{a}$ and $h_{b}$ in mind, 
let us examine the validity of (\ref{4.290}) 
in the light of the relevant experimental results 
of Wood \& Simpson (1984). 

When $h_{a}$ is kept constant in (\ref{4.290}), 
$U_{s}$ becomes a function of $h_{b}$ alone. 
By measuring the velocities of internal bores 
advancing into two stationary layers of fluid 
with a small density difference, 
Wood \& Simpson determined, 
for the values of $h_{a}=0.027$, $0.06$, and $0.12$, 
the dependence of $U_{s}$ on $h_{b}$. 
Their results are displayed in figure 3. 
\begin{figure}[tb]
 \centering
 \includegraphics{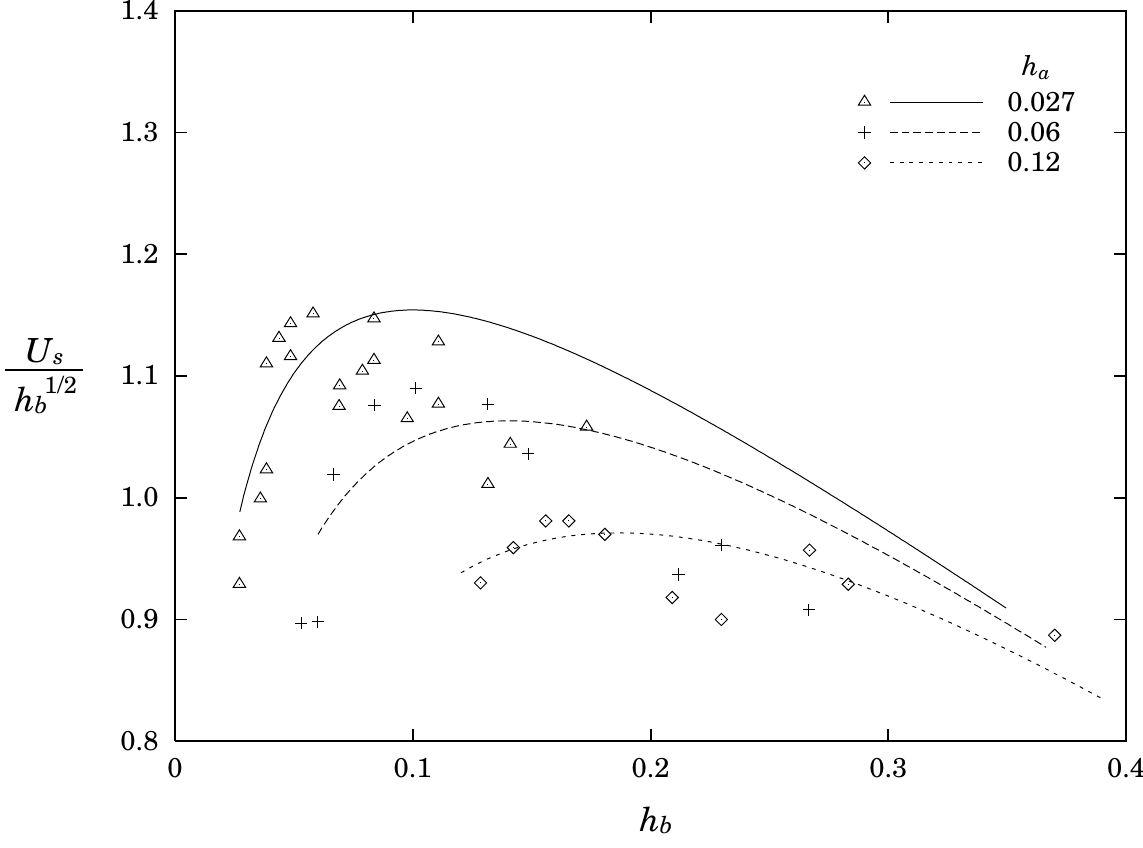}
 \caption[]{Dependence of $U_{s}$ on $h_{b}$
            for $h_{a}=0.027$, $0.06$, and $0.12$.
            The symbols show the experimental results
            of Wood \& Simpson (1984).
            The curves are obtained from (\ref{4.290}).}
\end{figure}
The theoretical curves determined from (\ref{4.290}) 
for the three values of $h_{a}$ are also shown for comparison: 
the curves are drawn for the ranges of $h_{b}$ 
where (\ref{4.350}) applies. 
The agreement between the theory and 
the experimental results is satisfactorily good 
in view of the uncertainty in determining $U_{s}$ and $h_{b}$, 
in particular the latter, from experiment. 

Finally, for comparison with (\ref{4.290}), 
we add an overview of the similar formulae 
which follow from the sets of shock conditions 
proposed in the past mentioned in \S\,1. 

We first consider the set of shock conditions 
proposed by Yih \& Guha (1955). 
This set of shock conditions is given by (\ref{4.90}) 
with its first shock condition replaced by 
\begin{equation}
  \label{4.360}
  \left[hv_{1}^{2}+\tfrac{1}{2} h^{2}+(1-\beta)h\eta\right]
 =\tfrac{1}{2}(1-\beta)(\eta_{a}+\eta_{b})(h_{a}-h_{b}), 
\end{equation}
where $\eta_{a}=\eta|_{x=x_{s}+0}$ and $\eta_{b}=\eta|_{x=x_{s}-0}$. 
Using this set of shock conditions, 
we can find, in place of (\ref{4.290}), the following formula: 
\begin{equation}
  \label{4.370}
  U_{s}=\left\{
        \frac{h_{b}(1-h_{b})(2-h_{a}-h_{b})(h_{a}+h_{b})}
             {2(h_{b}^2-2h_{a}h_{b}+2h_{a}-h_{a}^2)}
        \right\}^{\frac{1}{2}}.
\end{equation}

Of the two sets of shock conditions 
suggested by Chu \& Baddour (1977) and Wood \& Simpson (1984), 
one is obtained if the first shock condition 
in (\ref{4.90}) is replaced by 
\begin{equation}
  \label{4.380}
  \left[\tfrac{1}{2} v_{2}^{2}+\eta\right]=0.
\end{equation}
This set of shock conditions yields, 
in place of (\ref{4.290}), the following formula: 
\begin{equation}
  \label{4.390}
  U_{s}=\left\{
        \frac{h_{b}(1-h_{b})^2(h_{a}+h_{b})}
             {h_{b}^2-3h_{a}h_{b}+2h_{a}}
        \right\}^{\frac{1}{2}}.
\end{equation}

The other set of shock conditions suggested by 
Chu \& Baddour and Wood \& Simpson is obtained 
if the first shock condition in (\ref{4.90}) is replaced by 
\begin{equation}
  \label{4.400}
  \left[\tfrac{1}{2} v_{1}^{2}+h+(1-\beta)\eta\right]=0.
\end{equation}
From this set of shock conditions, 
we have, in place of (\ref{4.290}), the following formula: 
\begin{equation}
  \label{4.410}
  U_{s}=\left\{
        \frac{h_{b}^2(1-h_{b})(2-h_{a}-h_{b})}
             {h_{b}^2+h_{b}-3h_{a}h_{b}+h_{a}}
        \right\}^{\frac{1}{2}}.
\end{equation}

Wood \& Simpson (1984) found that (\ref{4.370}) and (\ref{4.390}) 
can predict the velocities of the bores in their experiments 
when the amplitudes of the bores are small enough; 
however, it was found at the same time that, 
when the amplitudes of the bores become large, 
the predictions from the formulae become unsatisfactory. 

Klemp et al.\ (1997) later showed 
that the velocities of the bores in the experiments of 
Wood \& Simpson can satisfactorily be predicted by (\ref{4.410}) 
irrespective of the amplitudes of the bores. 
However, as can readily be demonstrated, 
the set of shock conditions 
from which (\ref{4.410}) follows does not allow shocks 
to exist when the density in the upper layer is much smaller 
than that in the lower layer, i.e.\ when $\beta=1-0$; 
this result is evidently inconsistent with 
the theory of bores in classical hydraulics. 

Accordingly, we can conclude as follows: 
within the framework of 
the one-dimensional two-layer shallow-water equations, 
the set of shock conditions of Yih \& Guha 
and those of Chu \& Baddour and Wood \& Simpson 
all must be considered approximate ones 
that are adequate under specific circumstances. 

%% file: section5.tex
\section{Gravity currents}
In this section, we discuss three kinds of gravity currents 
within the framework of the one-dimensional two-layer 
shallow-water equations. 
For each kind of gravity current, 
we first derive the front conditions, 
i.e.\ the boundary conditions to be imposed 
at the front of the gravity current, 
from the four basic laws of the equations. 
We can then find from the front conditions 
a formula that gives the rate of advance of the gravity current 
as a function of its depth. 
After the energy condition and the evolutionary 
condition for the gravity current are discussed, 
this velocity formula is derived explicitly for a few special cases; 
the results are then compared with 
some empirical or theoretical formulae. 

\subsection{Gravity currents advancing along a no-slip lower boundary}
\subsubsection{Front conditions}
We consider again the physical situation in \S\,3. 
However, it is supposed now that 
the lower-layer fluid is advancing 
along the lower boundary as a gravity current. 
Behind the front of the gravity current, 
the fluid motion is taken to possess 
the properties described in \S\,3. 
Similarly, sufficiently ahead of the gravity current, 
the motion is assumed to have the same properties 
except that the whole depth of the channel 
is occupied there by the upper-layer fluid. 
Hence, to represent properly the motion in these regions, 
we use again the dimensionless coordinates $x$, $y$, $z$ 
and the dimensionless time $t$. 
We assume, without loss of generality, that the gravity current 
is advancing in the positive $x$-direction. 

The transition of the motion 
from the state sufficiently ahead of the gravity current 
to that behind the front of the current occurs 
inside a region containing the front. 
We call this region the frontal region. 
It is supposed that the motion inside the frontal region is, 
like the motion outside the region, 
two-dimensional for $|y|<\tfrac{1}{2}$. 

We now assume that, inside the frontal region, 
the interface between the layers 
meets the upper side of the lower vortex sheet, 
i.e.\ $z=\!\mbox{}+0$, at $x=x_{f}(t)$. 
Let $l_{f}$ be the scale 
of the dimensional distance across the frontal region; 
it is assumed that $l_{f}$ is very small compared with 
the length scale $L$ of the motion 
outside the region, i.e.\ $l_{f}/L\ll 1$. 
Then, in the dimensionless coordinate system, 
the frontal region may be idealized 
into a plane normal to the $x$-axis coincident with $x=x_{f}$. 
The lower layer is entirely absent 
ahead of the plane, i.e.\ for $x\ge x_{f}+0$, 
but has finite thickness behind it, i.e.\ for $x\le x_{f}-0$. 
Thus the front of the gravity current may be treated 
as a wall of fluid located at $x=x_{f}$. 

Let us next consider the motion 
in the infinite interval $x\ge x_{f}+0$. 
The fluid velocity in this interval 
is again expressed by (\ref{3.30}); 
however, since the lower layer is absent, 
the expression (\ref{3.50}) for the dimensionless velocity 
in the $x$-direction reduces to 
\begin{equation}
  \label{5.10}
  u=u_{2}(x,t),\qquad 0<z<1.
\end{equation}
The expression (\ref{3.60}) for the pressure distribution 
also reduces to 
\begin{equation}
  \label{5.20}
  p_{\ast}=\rho_{2}\beta gH\eta(x,t) +\rho_{2}gH(1-z), 
  \qquad |y|<\tfrac{1}{2},\quad 0 \le z \le 1. 
\end{equation}
Thus the motion in the infinite interval $x\ge x_{f}+0$ 
is specified in terms of the variables $u_{2}$ and $\eta$. 
These variables are governed by the equations 
\begin{equation}
  \label{5.30}
  \frac{\partial u_{2}}{\partial x}=0,
  \quad
  \frac{\partial \eta}{\partial x}
=-\frac{\partial u_{2}}{\partial t}-u_{2}\frac{\partial u_{2}}{\partial x}.
\end{equation}

On the other hand, the motion 
in the infinite interval $x\le x_{f}-0$ 
is specified in terms of the variables 
$u_{1}$, $u_{2}$, $h$, and $\eta$ introduced in \S\,3, 
and the variables are governed by 
the one-dimensional two-layer shallow-water equations (\ref{3.240}). 
Hence we need to solve (\ref{3.240}) simultaneously with (\ref{5.30}) 
to determine the motion for all $x\neq x_{f}$. 
To this end, however, it is necessary to know 
the boundary conditions concerning 
the variables $u_{1}$, $u_{2}$, $h$, and $\eta$ at $x=x_{f}-0$ 
and the variables $u_{2}$ and $\eta$ at $x=x_{f}+0$. 
Our aim is to formulate these boundary conditions 
which we call the front conditions. 
They can be obtained, 
in a way similar to the one to derive the shock conditions, 
from the four basic laws in \S\,3. 

Let us again focus our attention on the plane $y=0$, and 
consider the closed curve $\Gamma_{0}$ introduced in \S\,3. 
We assume that the arbitrary constants 
$x_{1}$ and $x_{2}$ are so chosen that $x_{1}>x_{f}>x_{2}$; 
the frontal region lies 
between $x=x_{1}$ and $x=x_{2}$, as shown in figure 4. 
\begin{figure}[tb]
 \centering
 \includegraphics{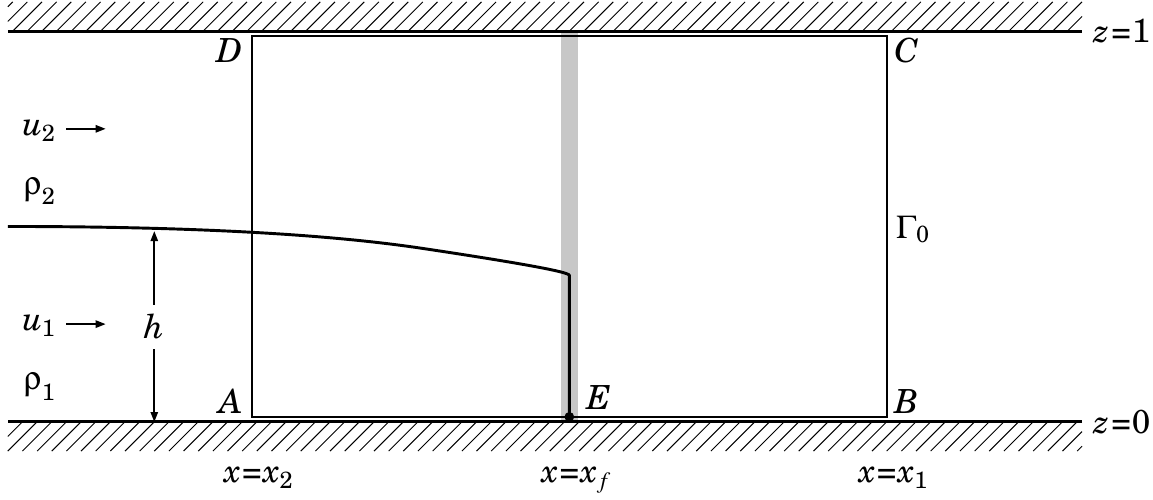}
 \caption[]{Position of the gravity current relative to $\Gamma_{0}$.
            The grey band shows the plane
            which represents the frontal region.
            The front of the current is contained in this plane,
            and the point of intersection of the interface
            and segment $AB$ is denoted by $E$.}
\end{figure}
In this figure, $E$ denotes the point of intersection 
of the interface and segment $AB$. 

Now let us assume that, 
despite the special choice of $x_{1}$ and $x_{2}$, 
the rate of change of the circulation around $\Gamma_{0}$ 
is still equal to the sum of 
the rate of advection of circulation across $\Gamma_{0}$ and 
the rate of generation of circulation due to baroclinicity. 
Then, from this conservation law of circulation 
for the interfacial vortex sheet, 
we can obtain one of the front conditions. 
To this end, we must first express the law mathematically. 

The rate of change of the circulation around $\Gamma_{0}$ 
can now be written as 
\begin{align}
  \label{5.40}
  \frac{\mathrm{d}}{\mathrm{d}t_{\ast}}\oint_{\Gamma_{0}}
       \boldsymbol{u}_{\ast}\cdot\boldsymbol{t}\:\mathrm{d}s_{\ast}
 &=
    \beta gH\frac{\mathrm{d}}{\mathrm{d}t}
    \left\{
       \int_{x_{2}}^{x_{f}-0}(u_{1}-u_{2})\:\mathrm{d}x
      +\int_{x_{f}+0}^{x_{1}}(u_{2}-u_{2})\:\mathrm{d}x
    \right\}
    \notag\\
 &\quad
   +\beta gH\frac{\mathrm{d}}{\mathrm{d}t}
       \int_{x_{f}-0}^{x_{f}+0}
       \frac{\left.\boldsymbol{u}_{\ast}\right|_{z=\mbox{}+0}
            -\left.\boldsymbol{u}_{\ast}\right|_{z=1-0}}
            {(\beta gH)^{\frac{1}{2}}}
       \cdot\boldsymbol{i}\:\mathrm{d}x.
\end{align}
However, it can easily be shown 
that the last term is negligible. Thus we have 
\begin{equation}
  \label{5.50}
  \frac{\mathrm{d}}{\mathrm{d}t_{\ast}}\oint_{\Gamma_{0}}
       \boldsymbol{u}_{\ast}\cdot\boldsymbol{t}\:\mathrm{d}s_{\ast}
 =\beta gH\frac{\mathrm{d}}{\mathrm{d}t}
  \int_{x_{2}}^{x_{f}-0}(u_{1}-u_{2})\:\mathrm{d}x.
\end{equation}

On the other hand, the rate of advection 
of circulation across $\Gamma_{0}$ can be written as 
\begin{align}
  \label{5.60}
    -\oint_{\Gamma_{0}}\omega_{n\ast}
        \boldsymbol{q}_{\ast}\cdot\boldsymbol{\nu}\:\mathrm{d}s_{\ast}
 &=
     \beta gH
     \left.(
        \tfrac{1}{2} u_{1}^{2}-\tfrac{1}{2} u_{2}^{2}
            )\right|_{x=x_{2}}
   -\beta gH
     \left.(
        \tfrac{1}{2} u_{2}^{2}-\tfrac{1}{2} u_{2}^{2}
            )\right|_{x=x_{1}}
     \notag\\
 &\quad
    -\beta gH
     \int_{x_{f}-0}^{x_{f}+0}
     \frac{\left.\omega_{n\ast}\boldsymbol{u}_{\ast}\right|_{z=1-0}
          -\left.\omega_{n\ast}\boldsymbol{u}_{\ast}\right|_{z=\mbox{}+0}}
          {\beta gH/L}
     \cdot\boldsymbol{k}\:\mathrm{d}x.
\end{align}
The second term on the right-hand side 
gives the rate of advection of circulation 
across segment $BC$, but vanishes identically. 
The last term of (\ref{5.60}) 
represents the contribution from 
the advection of circulation across $\Gamma_{0}$ 
occurring inside the frontal region. 
It seems reasonable to expect that this term also vanishes, 
for the upper and lower vortex sheets 
have been assumed never to separate from the boundaries. 
However, from the following discussion, 
it turns out that the term does not vanish 
contrary to this expectation. 

To see this, we first note the following fact: 
the lower vortex sheet is occupied 
by the upper-layer fluid for $x\ge x_{f}+0$ but 
by the lower-layer fluid for $x\le x_{f}-0$. 
This implies that, inside the frontal region, 
the upper-layer fluid in the lower vortex sheet 
is replaced by the lower-layer fluid 
as the gravity current advances. 
As the basis for our discussion, 
we introduce here a model of this process 
occurring inside the frontal region: 
the model is based on the observations of 
gravity currents by Simpson (1972). 

Let us consider the lower vortex sheet inside the frontal region. 
Ahead of the point of intersection of 
the interface and the upper side of the vortex sheet, 
i.e.\ ahead of point $E$ in figure 4, 
the vortex sheet is occupied by the upper-layer fluid. 
As the gravity current advances, 
$E$ also advances along the upper side of the vortex sheet. 
However, because of the influence 
of the no-slip condition at the lower boundary, 
the upper-layer fluid in the vortex sheet is left behind by $E$ 
beneath the following lower-layer fluid. 
In consequence, convective instability arises behind $E$: 
the upper-layer fluid left beneath the lower-layer fluid 
rises from the lower vortex sheet, 
penetrates the lower-layer fluid, and 
is absorbed into the interfacial vortex sheet; 
a fraction of the lower-layer fluid, on the other hand, 
is absorbed into the lower vortex sheet. 
The upper-layer fluid in the lower vortex sheet 
is replaced by the lower-layer fluid 
through this convective instability. 

Now let us study the influence of this process 
on the last term of (\ref{5.60}). 
We first need to recognize that, 
when the upper-layer fluid in the lower vortex sheet 
is left behind by point $E$, 
circulation also is left behind with the fluid. 
The amount of the circulation left behind by $E$ per unit time 
can be calculated from the following formula: 
\begin{equation}
  \label{5.70}
 -H\int_{0}^{\mbox{}+0}\left.\omega_{n\ast}\right|_{x=x_{f}}
  \left\{
    \left.\boldsymbol{q}_{\ast}\right|_{x=x_{f}}
    -(\beta gH)^{\frac{1}{2}}U_{f}\boldsymbol{i}
  \right\}
  \cdot\boldsymbol{i}\:\mathrm{d}z.
\end{equation}
Here $U_{f}(t)=\mathrm{d}x_{f}/\mathrm{d}t$ is the velocity of $E$, 
so that $\boldsymbol{q}_{\ast}-(\beta gH)^{\frac{1}{2}}U_{f}\boldsymbol{i}$ 
gives the dimensional surface velocity relative to $E$; 
the integral in (\ref{5.70}) is taken 
across the lower vortex sheet beneath $E$. 
Inside the lower vortex sheet, however, 
$\omega_{n\ast}$ can be approximated by 
\begin{equation}
  \label{5.80}
  \omega_{n\ast}
 =-\frac{1}{H}
   \frac{\partial}{\partial z}(\boldsymbol{u}_{\ast}\cdot\boldsymbol{i}).
\end{equation}
Substituting (\ref{5.80}) and $\boldsymbol{q}_{\ast}=\boldsymbol{u}_{\ast}$ 
into (\ref{5.70}) and carrying out the integration, we have 
\begin{equation}
  \label{5.90}
  \tfrac{1}{2}
  \left\{
    \left.\boldsymbol{u}_{\ast}\right|_{x=x_{f},z=\mbox{}+0}
    \cdot\boldsymbol{i}-(\beta gH)^{\frac{1}{2}}U_{f}
  \right\}^{2}
 -\tfrac{1}{2}
  \left\{
    \left.\boldsymbol{u}_{\ast}\right|_{x=x_{f},z=0}
    \cdot\boldsymbol{i}-(\beta gH)^{\frac{1}{2}}U_{f}
  \right\}^{2}.
\end{equation}
Note that $\boldsymbol{u}_{\ast}|_{x=x_{f},z=\mbox{}+0}$ 
and $\boldsymbol{u}_{\ast}|_{x=x_{f},z=0}$ 
are respectively the fluid velocities 
at $E$ and at the lower boundary. 
Since $E$ is a material point, 
these velocities satisfy the conditions 
\begin{equation}
  \label{5.100}
  \left.\boldsymbol{u}_{\ast}\right|_{x=x_{f},z=\mbox{}+0}
 =(\beta gH)^{\frac{1}{2}}U_{f}\boldsymbol{i},
  \quad
  \left.\boldsymbol{u}_{\ast}\right|_{x=x_{f},z=0}=0,
\end{equation}
where the latter follows from 
the no-slip condition at the lower boundary. 
Hence we see that the amount of the circulation 
left behind by $E$ per unit time is given by 
\begin{equation}
  \label{5.110}
 -\beta gH\tfrac{1}{2} U_{f}^{2}.
\end{equation}
It should be noted here that the circulation left behind by $E$, 
carried by the upper-layer fluid, rises from the lower vortex sheet 
and is finally absorbed into the interfacial vortex sheet. 
As a result, advection of circulation arises 
inside the frontal region across $\Gamma_{0}$. 
We can expect that the rate of 
this advection of circulation across $\Gamma_{0}$ 
is equal to (\ref{5.110}): 
\begin{equation}
  \label{5.120}
  \beta gH
  \int_{x_{f}-0}^{x_{f}+0}
  \frac{\left.\omega_{n\ast}\boldsymbol{u}_{\ast}\right|_{z=\mbox{}+0}}
       {\beta gH/L}
  \cdot\boldsymbol{k}\:\mathrm{d}x
=-\beta gH\tfrac{1}{2} U_{f}^{2}.
\end{equation}
Note that the lower-layer fluid absorbed 
into the lower vortex sheet does not contribute to 
the advection of circulation across $\Gamma_{0}$. 
This is a consequence of the fact that 
the fluid enters the vortex sheet from 
the region outside the vortex sheet where $\omega_{n\ast}=0$. 

On the other hand, no circulation leaks 
from the upper vortex sheet across $\Gamma_{0}$: 
\begin{equation}
  \label{5.130}
  \beta gH
  \int_{x_{f}-0}^{x_{f}+0}
  \frac{\left.\omega_{n\ast}\boldsymbol{u}_{\ast}\right|_{z=1-0}}
       {\beta gH/L}
  \cdot\boldsymbol{k}\:\mathrm{d}x=0.
\end{equation}
Substitution of (\ref{5.120}) and (\ref{5.130}) into 
the last term of (\ref{5.60}) leads to 
\begin{equation}
  \label{5.140}
 -\oint_{\Gamma_{0}}\omega_{n\ast}
    \boldsymbol{q}_{\ast}\cdot\boldsymbol{\nu}\:\mathrm{d}s_{\ast}
 =\beta gH
  \left.(
    \tfrac{1}{2} u_{1}^{2}-\tfrac{1}{2} u_{2}^{2}
         )\right|_{x=x_{2}}
 -\beta gH\tfrac{1}{2} U_{f}^{2}.
\end{equation}
This is the required expression for 
the rate of advection of circulation across $\Gamma_{0}$. 

Finally, we can write the rate of generation 
of circulation due to baroclinicity as 
\begin{equation}
  \label{5.150}
 -\oint_{\Gamma_{0}}\frac{1}{\rho_{\ast}}
              \nabla p_{\ast}\cdot\boldsymbol{t}\:\mathrm{d}s_{\ast} 
=-\left(\frac{1}{\rho_{1}}-\frac{1}{\rho_{2}}\right)
  \left\{
    \left.p_{\ast}\right|_{x=x_{f},z=\mbox{}+0}
   -\left.p_{\ast}\right|_{x=x_{2},z=h(x_{2},t)}
  \right\}.
\end{equation}
Here $p_{\ast}|_{x=x_{f},z=\mbox{}+0}$ denotes 
the dimensional pressure at point $E$ in figure 4. 
As explained below, this pressure can be expressed 
in terms of the pressure at point $B$ in figure 4. 

We first assume that Euler's equation of motion 
\begin{equation}
  \label{5.160}
  \frac{\partial \boldsymbol{u}_{\ast}}{\partial t_{\ast}}
 +\nabla(\tfrac{1}{2}\boldsymbol{u}_{\ast}\cdot\boldsymbol{u}_{\ast})
 -\boldsymbol{u}_{\ast}\times \boldsymbol{\omega}_{\ast}
=-\frac{1}{\rho_{\ast}}\nabla p_{\ast}-g\boldsymbol{k}
\end{equation}
is valid on segment $EB$ in figure 4, 
and take the line integral along $EB$ of the equation. 
Since $EB$ is a streamline, 
the integral of the third term vanishes. 
In addition, since $\boldsymbol{k}$ is normal to $EB$, 
the integral of the last term vanishes as well. 
Thus we obtain 
\begin{align}
  \label{5.170}
  &
  \beta gH
  \int_{x_{f}}^{x_{1}}\frac{\partial}{\partial t}
  \left\{
  \frac{\left.\boldsymbol{u}_{\ast}\right|_{z=\mbox{}+0}}
       {(\beta gH)^{\frac{1}{2}}}
  \right\}\cdot\boldsymbol{i}\:\mathrm{d}x
  +
  \beta gH
  \left(
  \left.\tfrac{1}{2} u_{2}^{2}\right|_{x=x_{1}}-\tfrac{1}{2} U_{f}^{2}
  \right)
  \notag\\
&\qquad\qquad
 =
 -\frac{1}{\rho_{2}}
  \left(
  \left.p_{\ast}\right|_{x=x_{1},z=\mbox{}+0}
 -\left.p_{\ast}\right|_{x=x_{f},z=\mbox{}+0}
  \right).
\end{align}
Here we have used the former condition of (\ref{5.100}) and 
$\boldsymbol{u}_{\ast}|_{x=x_{1},z=\mbox{}+0}
=(\beta gH)^{\frac{1}{2}}u_{2}|_{x=x_{1}}\boldsymbol{i}$. 
The former of (\ref{5.100}) also enables us to rewrite 
the first term of (\ref{5.170}) as 
\begin{equation}
  \label{5.180}
  \beta gH \frac{\mathrm{d}}{\mathrm{d}t}\int_{x_{f}}^{x_{f}+0}
  \frac{\left.\boldsymbol{u}_{\ast}\right|_{z=\mbox{}+0}}
       {(\beta gH)^{\frac{1}{2}}}
  \cdot\boldsymbol{i}\:\mathrm{d}x
 +\beta gH 
  \frac{\mathrm{d}}{\mathrm{d}t}\int_{x_{f}+0}^{x_{1}}u_{2}\:\mathrm{d}x
 +\beta gH U_{f}^{2},
\end{equation}
where $\boldsymbol{u}_{\ast}|_{z=\mbox{}+0}
=(\beta gH)^{\frac{1}{2}}u_{2}\boldsymbol{i}$ 
has been used for $x\ge x_{f}+0$. 
However, like the last term of (\ref{5.40}), 
the first term of (\ref{5.180}) is negligible. 
Thus it is seen that (\ref{5.170}) yields 
\begin{equation}
  \label{5.190}
  \left.p_{\ast}\right|_{x=x_{f},z=\mbox{}+0}
 =\left.p_{\ast}\right|_{x=x_{1},z=\mbox{}+0}
 +\rho_{2}\beta gH
  \left(
  \left.\tfrac{1}{2} u_{2}^{2}\right|_{x=x_{1}}+\tfrac{1}{2} U_{f}^{2}
 +\frac{\mathrm{d}}{\mathrm{d}t}\int_{x_{f}+0}^{x_{1}}u_{2}\:\mathrm{d}x
  \right).
\end{equation}
This formula expresses the pressure at $E$ 
in terms of that at $B$. 

Substituting (\ref{5.190}) into (\ref{5.150}) 
and then using (\ref{3.60}) and (\ref{5.20}), 
we obtain the following expression for 
the rate of generation of circulation due to baroclinicity: 
\begin{align}
  \label{5.200}
 -\oint_{\Gamma_{0}}\frac{1}{\rho_{\ast}}
              \nabla p_{\ast}\cdot\boldsymbol{t}\:\mathrm{d}s_{\ast} 
 &=
    \beta gH
    \left\{
    \left.(h-\beta\eta)\right|_{x=x_{2}}
   +\left.\beta\eta\right|_{x=x_{1}}
    \right\}
    \notag\\
 &\quad
   +\beta^{2} gH
    \left(
    \left.\tfrac{1}{2} u_{2}^{2}\right|_{x=x_{1}}+\tfrac{1}{2} U_{f}^{2}
   +\frac{\mathrm{d}}{\mathrm{d}t}\int_{x_{f}+0}^{x_{1}}u_{2}\:\mathrm{d}x
    \right).
\end{align}
Hence, from (\ref{5.50}), (\ref{5.140}), and (\ref{5.200}), 
we see that the conservation law of circulation 
for the interfacial vortex sheet is expressed by 
\begin{align}
  \label{5.210}
  \frac{\mathrm{d}}{\mathrm{d}t}
  \int_{x_{2}}^{x_{f}-0}(u_{1}-u_{2})\:\mathrm{d}x
 &=
  \left.(
  \tfrac{1}{2} u_{1}^{2}-\tfrac{1}{2} u_{2}^{2}+h-\beta\eta
         )\right|_{x=x_{2}}
 +\left.\beta\eta\right|_{x=x_{1}}
 -\tfrac{1}{2} U_{f}^{2}
  \notag\\
 &\quad
 +\beta
  \left(
  \left.\tfrac{1}{2} u_{2}^{2}\right|_{x=x_{1}}+\tfrac{1}{2} U_{f}^{2}
 +\frac{\mathrm{d}}{\mathrm{d}t}\int_{x_{f}+0}^{x_{1}}u_{2}\:\mathrm{d}x
  \right).
\end{align}
Note that the third term on the right-hand side of (\ref{5.210}) 
represents the influence of the lower vortex sheet 
on the balance of circulation for the interfacial vortex sheet. 

Now, applying the procedure described by Whitham (1974, \S\,5.8)
to (\ref{5.210}), we find 
\begin{align}
  \label{5.220}
  \left.U_{f}(u_{1}-u_{2})\right|_{x=x_{f}-0}
 &=
  \left.(
  \tfrac{1}{2} u_{1}^{2}-\tfrac{1}{2} u_{2}^{2}+h-\beta\eta
         )\right|_{x=x_{f}-0}
 +\left.\beta\eta\right|_{x=x_{f}+0}
 -\tfrac{1}{2} U_{f}^{2}\makebox[1em]{}
  \notag\\
 &\quad
 +\left.\beta\tfrac{1}{2}(u_{2}- U_{f})^{2}\right|_{x=x_{f}+0}.
\end{align}
This is the front condition corresponding to the conservation law 
of circulation for the interfacial vortex sheet. 
The third term on the right-hand side of (\ref{5.220}) 
stems from that of (\ref{5.210}), 
so that the influence of the lower vortex sheet 
on the balance of circulation for the interfacial vortex sheet 
is represented by this term. 

In contrast, the influence of the lower vortex sheet 
on the balance of mass and that of momentum 
can readily be shown to be negligible. 
Hence the remaining front conditions can be obtained 
much more easily than (\ref{5.220}): 
from the conservation law of momentum 
for the upper and lower layers together, we obtain 
\begin{align}
  \label{5.230}
  &
  \left.U_{f}
  \left\{hu_{1}+(1-\beta)(1-h)u_{2}\right\}
  \right|_{x=x_{f}-0}
 -\left.U_{f}(1-\beta)u_{2}\right|_{x=x_{f}+0}
  \notag\\
&\quad
  =
  \left.\left\{
    hu_{1}^{2}+(1-\beta)(1-h)u_{2}^{2}+\tfrac{1}{2} h^{2}+(1-\beta)\eta
  \right\}\right|_{x=x_{f}-0}
  \notag\\
&\qquad
 -\left.\left\{
   (1-\beta)u_{2}^{2}+(1-\beta)\eta
 \right\}\right|_{x=x_{f}+0};
\end{align}
from the conservation law of mass for the lower layer, 
\begin{equation}
  \label{5.240}
  \left.U_{f}h\right|_{x=x_{f}-0}=\left.hu_{1}\right|_{x=x_{f}-0};
\end{equation}
and from the conservation law of mass for the upper layer, 
\begin{align}
  \label{5.250}
  &
  \left.U_{f}(1-\beta)(1-h)\right|_{x=x_{f}-0}
 -U_{f}(1-\beta)
  \notag\\
 &\quad
  =
  \left.(1-\beta)(1-h)u_{2}\right|_{x=x_{f}-0}
 -\left.(1-\beta)u_{2}\right|_{x=x_{f}+0}.
\end{align}
The conditions (\ref{5.220})--(\ref{5.250}) form 
the set of front conditions for the gravity current. 
In (\ref{5.220})--(\ref{5.250}), $U_{f}$ may be regarded 
as the rate of advance of the gravity current. 

It is convenient to express the front conditions 
in terms of the relative velocities 
\begin{equation}
  \label{5.260}
  v_{1}=u_{1}-U_{f},\quad v_{2}=u_{2}-U_{f}.
\end{equation}
Substituting (\ref{5.260}) into (\ref{5.220}), we have 
\begin{equation}
  \label{5.270}
  \left.(
  \tfrac{1}{2} v_{1}^{2}-\tfrac{1}{2} v_{2}^{2}+h-\beta\eta
         )\right|_{x=x_{f}-0}
 +\left.\beta\eta\right|_{x=x_{f}+0}
 +\left.\beta\tfrac{1}{2} v_{2}^{2}\right|_{x=x_{f}+0}
 =\tfrac{1}{2} U_{f}^{2}.
\end{equation}
Here the term on the right-hand side represents 
the influence of the lower vortex sheet. 
Also, from (\ref{5.230})--(\ref{5.250}), we obtain 
\begin{equation}
  \label{5.280}
  \left.
  \begin{array}{c}
  \displaystyle
  \left.hv_{1}\right|_{x=x_{f}-0}=0,
  \quad
  \left.(1-\beta)(1-h)v_{2}\right|_{x=x_{f}-0}
 -\left.(1-\beta)v_{2}\right|_{x=x_{f}+0}=0,\\[6pt]
  \displaystyle
  \left.\left\{
    hv_{1}^{2}+(1-\beta)(1-h)v_{2}^{2}+\tfrac{1}{2} h^{2}+(1-\beta)\eta
  \right\}\right|_{x=x_{f}-0}\\[6pt]
  \displaystyle
  \mbox{}
 -\left.\left\{
   (1-\beta)v_{2}^{2}+(1-\beta)\eta
 \right\}\right|_{x=x_{f}+0}=0.\makebox[0.86em]{}
  \end{array}
  \right\}
\end{equation}

In conclusion, it must be stressed that 
the above front conditions have been obtained on the assumption 
that the gravity current is advancing 
relative to the lower boundary: 
this assumption was explicitly used 
when we evaluated the influence of the lower vortex sheet 
on the balance of circulation for the interfacial vortex sheet. 
Hence it follows that the above front conditions, 
or (\ref{5.220}) and (\ref{5.270}) to be more exact, 
are applicable only to a gravity current 
advancing relative to a no-slip lower boundary. 

\subsubsection{Energy condition and evolutionary condition}
The gravity current in \S\,5.1.1 satisfies 
two additional conditions corresponding to 
the energy condition and the evolutionary condition 
stated in \S\,4 about internal bores. 
Our next aim is to formulate these additional conditions 
for the gravity current. 

The energy condition for the gravity current 
can be expressed most concisely 
in terms of the relative velocity $v_{2}$. 
We first note that the second front condition 
in (\ref{5.280}) yields 
\begin{equation}
  \label{5.290}
  \left.(1-\beta)(1-h)v_{2}\right|_{x=x_{f}-0}
 =\left.(1-\beta)v_{2}\right|_{x=x_{f}+0}
 =I_{f}(t),
\end{equation}
where $I_{f}$ represents the rate of total advection of mass 
across the frontal region of the gravity current. 
We next introduce the following quantity: 
\begin{equation}
  \label{5.300}
  \Phi_{f}(t)
 =I_{f}
  \left\{
  \left.(\tfrac{1}{2} v_{2}^{2}+\eta)\right|_{x=x_{f}-0}
 -\left.(\tfrac{1}{2} v_{2}^{2}+\eta)\right|_{x=x_{f}+0}
  \right\}.
\end{equation}
In terms of this quantity, the rate of dissipation 
of mechanical energy inside the frontal region 
is given by $\rho_{1}(\beta gH)^{\frac{3}{2}}HW\Phi_{f}$. 
Thus we see that the gravity current satisfies 
\begin{equation}
  \label{5.310}
  \Phi_{f}\ge 0.
\end{equation}
This is the energy condition for the gravity current. 

Here, as in \S\,4.2, a remark needs to be made 
on the distribution of mechanical energy dissipation 
inside the frontal region. 
The dissipation of mechanical energy may arise, 
on account of turbulence, 
both in the gravity current and in the ambient fluid; 
however, we cannot predict the distribution of the dissipation, 
for it is impossible to predict 
the rate of transfer of mechanical energy 
between the gravity current and the ambient fluid. 

Let us next consider the evolutionary condition 
for the gravity current. 
This condition can be stated as follows: 
there are, at any instant, $N-1$ characteristics 
leaving the front of the gravity current and $M-N$ characteristics 
reaching the front or moving with the front. 
Here $N$ is the number of the front conditions, 
and $M$ the number of the variables in the front conditions. 
Apparently, $N$ is four. 
On the other hand, $M$ is seven: 
$u_{1}$, $u_{2}$, $h$, and $\eta$ at $x=x_{f}-0$, 
$u_{2}$ and $\eta$ at $x=x_{f}+0$, and $U_{f}$. 

To express mathematically 
the evolutionary condition for the gravity current, 
we now recall that the fluid motion is governed by 
the one-dimensional two-layer shallow-water 
equations (\ref{3.240}) for $x\le x_{f}-0$. 
As we have seen in \S\,4.3, the system (\ref{3.240}) 
has the four families of characteristics 
$C_{+}$, $C_{-}$, $C_{vol}$, and $C_{\eta}$. 
For $x\ge x_{f}+0$, however, the motion is governed by 
the equations (\ref{5.30}). 
It can easily be seen that the system (\ref{5.30}) 
has the two families of characteristics $C_{vol}$ and $C_{\eta}$. 
Accordingly, a discussion similar to that in \S\,4.3 
leads us to the conclusion that the evolutionary condition 
is equivalent to 
\begin{equation}
  \label{5.320}
  \lambda_{-}-U_{f}<0\le\lambda_{+}-U_{f}
  \quad \mbox{at} \quad x=x_{f}-0,
\end{equation}
where $\lambda_{+}-U_{f}$ and $\lambda_{-}-U_{f}$ denote 
the characteristic velocities of $C_{+}$ and $C_{-}$ 
relative to the front of the gravity current. 
We can calculate the relative characteristic velocities 
again from (\ref{4.240}) and (\ref{4.250}) 
by replacing $U_{s}$ in (\ref{4.240}) with $U_{f}$: 
we must, of course, take 
$v_{1}$ and $v_{2}$ in the formulae 
as the relative velocities defined by (\ref{5.260}). 

\subsubsection{Velocity formula}
The front conditions (\ref{5.220})--(\ref{5.250}) 
for the gravity current in \S\,5.1.1 
may be considered a system of four algebraic equations 
for the following seven variables: 
$u_{1}$, $u_{2}$, $h$, and $\eta$ at $x=x_{f}-0$, 
$u_{2}$ and $\eta$ at $x=x_{f}+0$, and $U_{f}$. 
Hence, if we are given the values of 
the two variables at $x=x_{f}+0$, 
$U_{f}$ can be determined from the front conditions 
as a function of one of the four variables at $x=x_{f}-0$. 
In the following, we derive specifically a formula 
that gives $U_{f}$ as a function of $h$ at $x=x_{f}-0$, 
and then compare it with some empirical formulae 
so as to confirm the validity of the front conditions. 

To derive this velocity formula, we first need 
to determine $u_{2}$ and $\eta$ for $x\ge x_{f}+0$. 
It is seen from (\ref{5.30}) that these variables 
must take constant values for $x\ge x_{f}+0$. 
Thus, if the constant value of $\eta$ is put equal to zero 
without loss of generality, we have 
\begin{equation}
  \label{5.330}
  u_{2}=U_{\infty}, \quad\eta=0, \qquad x\ge x_{f}+0,
\end{equation}
where $U_{\infty}$ denotes the value of $u_{2}$ at $x=\infty$. 

Consider now the front conditions (\ref{5.270}) and (\ref{5.280}) 
expressed in terms of the relative velocities (\ref{5.260}). 
The parameter $\beta$ in the conditions 
can take any value between $\!\mbox{}+0$ and $1-0$, 
but we assume particularly 
that the density in the gravity current 
is almost equal to the ambient density, i.e.\ $\beta=\!\mbox{}+0$. 
Then (\ref{5.270}) reduces to 
\begin{equation}
  \label{5.340}
  \left.(
  \tfrac{1}{2} v_{1}^{2}-\tfrac{1}{2} v_{2}^{2}+h
         )\right|_{x=x_{f}-0}
 =\tfrac{1}{2} U_{f}^{2}.
\end{equation}
The conditions (\ref{5.280}) also reduce to 
\begin{equation}
  \label{5.350}
  \left.
  \begin{array}{c}
  \displaystyle
  \left.hv_{1}\right|_{x=x_{f}-0}=0,
  \quad
  \left.(1-h)v_{2}\right|_{x=x_{f}-0}
 =U_{\infty}-U_{f},\\[6pt]
  \displaystyle
  \left.\left\{
    hv_{1}^{2}+(1-h)v_{2}^{2}+\tfrac{1}{2} h^{2}+\eta
  \right\}\right|_{x=x_{f}-0}
 =(U_{\infty}-U_{f})^{2}.
  \end{array}
  \right\}
\end{equation}
Here we have used the following conditions 
obtained from (\ref{5.260}) and (\ref{5.330}): 
\begin{equation}
  \label{5.360}
  \left. v_{2}\right|_{x=x_{f}+0}=U_{\infty}-U_{f},\quad
  \left. \eta\right|_{x=x_{f}+0}=0.
\end{equation}

If (\ref{5.340}) and the first two front conditions 
in (\ref{5.350}) are used, we can determine $U_{f}$ 
as a function of $h$ at $x=x_{f}-0$. 
Introducing the notation 
\begin{equation}
  \label{5.370}
  h_{b}=\left.h\right|_{x=x_{f}-0},
\end{equation}
we can express the result as follows: 
\begin{equation}
  \label{5.380}
  U_{f}
 =\frac{U_{\infty}}{1+(1-h_{b})^{2}}
 +\frac{(1-h_{b})
        \left\{
        2h_{b}+2h_{b}(1-h_{b})^{2}-U_{\infty}^{2}
        \right\}^{\frac{1}{2}}}
       {1+(1-h_{b})^{2}}.
\end{equation}
This is the required velocity formula 
under the condition $\beta=\!\mbox{}+0$. 

It deserves attention 
that (\ref{5.380}) has been obtained from (\ref{5.340}) 
and the first two front conditions in (\ref{5.350}). 
We see from this fact that (\ref{5.380}) is based on 
the conservation laws of mass and of circulation; 
it is independent of the conservation law of momentum. 

We should also note 
that (\ref{5.380}) is valid only when $U_{f}>0$. 
This can be seen from the fact 
stated at the end of \S\,5.1.1: 
the front condition (\ref{5.270}), 
from which (\ref{5.340}) follows, 
is valid only when the gravity current is advancing 
relative to the lower boundary. 

Now let the ambient fluid be stationary 
sufficiently ahead of the gravity current. 
The velocity formula (\ref{5.380}) then reduces to 
\begin{equation}
  \label{5.390}
  U_{f}
 =(1-h_{b})
  \left\{\frac{2h_{b}}{1+(1-h_{b})^{2}}\right\}^{\frac{1}{2}},
\end{equation}
since $U_{\infty}=0$. 
For a while, we concentrate on the discussion of (\ref{5.390}). 

The evolutionary condition (\ref{5.320}) places 
the following restriction on $h_{b}$ in (\ref{5.390}): 
\begin{equation}
  \label{5.410}
  \!\mbox{}+0\le h_{b}\le 0.404. 
\end{equation}
We can also verify that 
the energy condition (\ref{5.310}) holds 
over the range (\ref{5.410}). 

Figure 5 shows a graph of (\ref{5.390}): 
it is drawn for the range (\ref{5.410}). 
\begin{figure}[tb]
 \centering
 \includegraphics{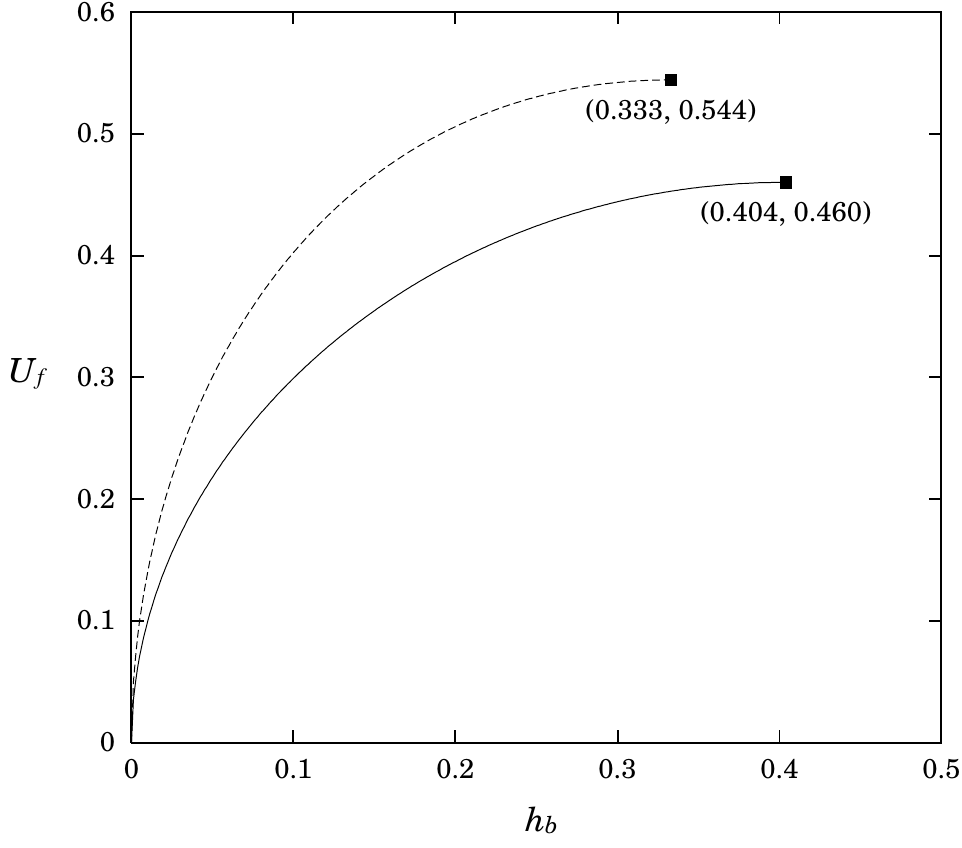}
 \caption[]{The solid line shows a graph of (\ref{5.390}).
            Also shown for comparison is a graph of (\ref{5.600}):
            this formula applies when the no-slip condition
            at the lower boundary is relaxed.}
\end{figure}
As we can see from the graph, 
$U_{f}$ takes the maximum value when $h_{b}=0.404$. 
This maximum value is 
\begin{equation}
  \label{5.420}
  U_{f}=0.460.
\end{equation}
Note also that, when $h_{b}=0.404$, 
the gravity current is marginally evolutionary: 
\begin{equation}
  \label{5.430}
  \lambda_{-}<U_{f}=\lambda_{+}
  \quad \mbox{at} \quad x=x_{f}-0.
\end{equation}
Klemp et al.\ (1994) showed 
that a marginally evolutionary gravity current is formed 
in what is called a lock-exchange experiment. 
Thus it is expected 
that the rate of advance of the gravity current formed 
in a lock-exchange experiment is given by (\ref{5.420}). 

An empirical formula that gives the rate of advance 
of the gravity current formed in a lock-exchange experiment 
can be found in the monograph of Yih (1965, p.~136): 
\begin{equation}
  \label{5.440}
  U_{f}=0.67(2-\beta)^{-\frac{1}{2}}.
\end{equation}
Though (\ref{5.440}) contains the parameter $\beta$, 
it was in fact obtained for very small values of $\beta$. 
Hence we may safely put $\beta=\!\mbox{}+0$ in (\ref{5.440}). 
Then we have 
\begin{equation}
  \label{5.450}
  U_{f}=0.47. 
\end{equation}
As has been expected, (\ref{5.450}) agrees well 
with (\ref{5.420}). 

On the other hand, as $h_{b}\rightarrow\!\mbox{}+0$, 
(\ref{5.390}) asymptotically approaches 
\begin{equation}
  \label{5.460}
  U_{f}=h_{b}^{\frac{1}{2}}.
\end{equation}
This asymptotic behaviour of (\ref{5.390}) coincides 
with that of the empirical formula 
\begin{equation}
  \label{5.470}
  U_{f}
 =\left\{
        \frac{h_{b}(1-h_{b})(2-h_{b})}{2(1+h_{b})}
  \right\}^{\frac{1}{2}}
\end{equation}
proposed by Rottman \& Simpson (1983) and 
shown to be adequate when $h_{b}$ is small. 

Having considered the special case in which $U_{\infty}=0$, 
we now return to the discussion of 
the original velocity formula (\ref{5.380}) 
to examine the dependence of $U_{f}$ on $U_{\infty}$. 

We should first recall that $U_{f}$ given by (\ref{5.380}) 
must satisfy the underlying assumption $U_{f}>0$. 
It is seen from (\ref{5.380}) that, 
for $U_{f}>0$ to be fulfilled by some value of $h_{b}$, 
\begin{equation}
  \label{5.490}
  -({\textstyle\frac{2}{3}})^{\frac{3}{2}}
  <U_{\infty}<2^{\frac{1}{2}}
\end{equation}
needs to be satisfied. 
We may interpret (\ref{5.490}) as a necessary condition 
for the existence of a gravity current 
advancing relative to a no-slip lower boundary. 

When $h_{b}$ is assumed to be constant in (\ref{5.380}), 
$U_{f}$ may be considered a function of $U_{\infty}$ alone. 
If $U_{\infty}^{2}\ll 2h_{b}\{1+(1-h_{b})^{2}\}$ in addition, 
then (\ref{5.380}) is approximated by 
\begin{equation}
  \label{5.500}
  U_{f}
 =\frac{U_{\infty}}{1+(1-h_{b})^{2}}
 +(1-h_{b})
  \left\{\frac{2h_{b}}{1+(1-h_{b})^{2}}\right\}^{\frac{1}{2}}.
\end{equation}
It is of interest to compare 
this approximate velocity formula and the empirical formula 
obtained by Simpson \& Britter (1980) stated below. 

Simpson \& Britter examined 
the dependence of $U_{f}$ on $U_{\infty}$ by experiment. 
They tried to express $U_{f}$, 
as a function of $U_{\infty}$, in the form 
\begin{equation}
  \label{5.510}
  U_{f}
 =aU_{\infty}+bh_{b}^{\frac{1}{2}},
\end{equation}
and found that $(a,b)=(0.62,0.91)$ for $0.15<h_{b}<0.25$. 
However, if we compare (\ref{5.500}) and (\ref{5.510}), 
it is seen that $a$ and $b$ in (\ref{5.510}) 
can be written as 
\begin{equation}
  \label{5.520}
  a=\frac{1}{1+(1-h_{b})^{2}},\quad
  b=(1-h_{b})
    \left\{\frac{2}{1+(1-h_{b})^{2}}\right\}^{\frac{1}{2}}.
\end{equation}
When $h_{b}=0.2$, for example, 
(\ref{5.520}) predicts that $(a,b)=(0.61,0.88)$. 
This prediction is in good agreement 
with the above result of Simpson \& Britter. 

Finally, we wish to discuss briefly 
the case in which the density in the gravity current 
is much larger than the ambient density, 
i.e.\ the case in which $\beta=1-0$. 

To this end, we need to return to the consideration of  
the front conditions (\ref{5.270}) and (\ref{5.280}). 
It can readily be seen from (\ref{5.270}) and (\ref{5.280}) 
that, when $\beta=1-0$, $U_{f}$ cannot be determined 
as a function of $h$ at $x=x_{f}-0$. 
Instead, the conditions require that 
\begin{equation}
  \label{5.550}
  \left.h\right|_{x=x_{f}-0}=\!\mbox{}+0.
\end{equation}
Thus, when $\beta=1-0$, the front of the gravity current 
cannot take the form of a wall of fluid. 
This is a fact well known in classical hydraulics 
(see Whitham 1974, \S\,13.10). 

\subsection{Gravity currents advancing 
            along a lower boundary with slip}
\subsubsection{Front conditions}
Let us consider again the gravity current in \S\,5.1.1. 
However, we now assume 
that the upper- and lower-layer fluids 
are allowed to slip at the lower boundary. 
Accordingly, the vortex sheet 
on the lower boundary is now entirely absent. 

The absence of the lower vortex sheet affects 
the front condition corresponding to the conservation law 
of circulation for the interfacial vortex sheet: 
the term representing 
the influence of the vortex sheet vanishes. 
For example, (\ref{5.270}) now becomes 
\begin{equation}
  \label{5.570}
  \left.(
  \tfrac{1}{2} v_{1}^{2}-\tfrac{1}{2} v_{2}^{2}+h-\beta\eta
         )\right|_{x=x_{f}-0}
 +\left.\beta\eta\right|_{x=x_{f}+0}
 +\left.\beta\tfrac{1}{2} v_{2}^{2}\right|_{x=x_{f}+0}=0.
\end{equation}
In contrast, 
the remaining front conditions (\ref{5.280}) are unaltered. 

\subsubsection{Energy condition and evolutionary condition}
The energy condition and the evolutionary condition 
for the gravity current in \S\,5.1.1 are unaffected 
by the absence of the lower vortex sheet. 
Hence the energy condition 
can be expressed by (\ref{5.310}) as before, 
and the evolutionary condition by (\ref{5.320}). 

\subsubsection{Velocity formula}
Now let us derive the velocity formula 
from the front conditions stated in \S\,5.2.1. 
The derivation is to be carried out 
for two special cases separately. 

The first case is the one 
in which the density in the gravity current 
is almost equal to the ambient density, 
i.e.\ the case in which $\beta=\!\mbox{}+0$. 
In this case, as explained in \S\,5.1.3, 
the conditions (\ref{5.280}) reduce to (\ref{5.350}). 
On the other hand, (\ref{5.570}) becomes 
\begin{equation}
  \label{5.580}
  \left.(
  \tfrac{1}{2} v_{1}^{2}-\tfrac{1}{2} v_{2}^{2}+h
         )\right|_{x=x_{f}-0}=0.
\end{equation}
From (\ref{5.580}) and 
the first two front conditions in (\ref{5.350}), we obtain 
\begin{equation}
  \label{5.590}
  U_{f}
 =U_{\infty}+(1-h_{b})(2h_{b})^{\frac{1}{2}}.
\end{equation}

Note that (\ref{5.590}) expresses $U_{f}$ 
as the sum of $U_{\infty}$ and a term independent of $U_{\infty}$. 
This implies that, in the absence of the lower vortex sheet, 
the rate of advance of the gravity current 
relative to the ambient fluid is independent of 
the velocity of the ambient fluid. 

Let us now suppose that $U_{\infty}=0$. 
Then (\ref{5.590}) reduces to 
\begin{equation}
  \label{5.600}
  U_{f}
 =(1-h_{b})(2h_{b})^{\frac{1}{2}}.
\end{equation}
The evolutionary condition (\ref{5.320}) requires 
that $h_{b}$ in (\ref{5.600}) should lie in the range 
\begin{equation}
  \label{5.610}
  \!\mbox{}+0\le h_{b}\le{\textstyle\frac{1}{3}}.
\end{equation}
The energy condition (\ref{5.310}) 
is automatically satisfied for this range of $h_{b}$. 

The formula (\ref{5.600}) should be compared 
with (\ref{5.390}) derived in \S\,5.1.3 
in the presence of the lower vortex sheet. 
To this end, we have included in figure 5 
a graph of (\ref{5.600}) for the range (\ref{5.610}). 
Comparing the graph with that of (\ref{5.390}), 
we see that, for any value of $h_{b}$ 
in the range (\ref{5.610}), 
(\ref{5.600}) gives a larger value of $U_{f}$ than (\ref{5.390}).
It is also seen from the graph that, 
when $h_{b}=\frac{1}{3}$, (\ref{5.600}) gives 
the following maximum value of $U_{f}$: 
\begin{equation}
  \label{5.620}
  U_{f}=({\textstyle\frac{2}{3}})^{\frac{3}{2}}=0.544.
\end{equation}
This is again larger than the maximum value (\ref{5.420}) 
obtained from (\ref{5.390}). 
These results allow us to say 
that the gravity current is retarded 
by friction at the lower boundary. 
It must be emphasized, however, 
that the retardation of the gravity current is not directly caused 
by the frictional force exerted by the lower boundary; 
the retardation is, in fact, due to the advection of circulation 
from the lower vortex sheet elucidated in \S\,5.1.1. 

It is also of interest to compare (\ref{5.600}) 
with (\ref{4.290}) derived in \S\,4.4 for an internal bore 
advancing into two stationary layers of fluid. 
Comparing the formulae, we observe that, 
if $U_{s}$ in (\ref{4.290}) is identified with $U_{f}$, 
(\ref{5.600}) is obtained from (\ref{4.290}) 
when $h_{a}\ll h_{b}$. 
Hence the gravity current may be regarded as 
an extreme form of the internal bore considered in \S\,4.4, 
though this is not allowed when the lower vortex sheet is present. 

Now let us turn to the consideration of the second case. 
This case is the one in which 
the ambient fluid is much deeper than the gravity current, 
i.e.\ the case in which $h\ll 1$ for $x\le x_{f}-0$. 
Though the value of $\beta$ is left unspecified, 
it is assumed that $1-\beta$ is not very small. 
To derive the velocity formula, 
we first introduce the approximation 
\begin{equation}
  \label{5.630}
  \left.(1-h)\right|_{x=x_{f}-0}=1.
\end{equation}
If this approximation and (\ref{5.360}) are used, 
(\ref{5.280}) can be simplified to 
\begin{equation}
  \label{5.640}
  \left.
  \begin{array}{c}
  \displaystyle
  \left.hv_{1}\right|_{x=x_{f}-0}=0,
  \quad
  \left.v_{2}\right|_{x=x_{f}-0}=U_{\infty}-U_{f},\\[6pt]
  \displaystyle
  \left.\left\{
    hv_{1}^{2}+(1-\beta)v_{2}^{2}+\tfrac{1}{2} h^{2}+(1-\beta)\eta
  \right\}\right|_{x=x_{f}-0}
 =(1-\beta)(U_{\infty}-U_{f})^{2}.
  \end{array}
  \right\}
\end{equation}
On the other hand, substitution of (\ref{5.360}) 
into (\ref{5.570}) leads to 
\begin{equation}
  \label{5.650}
  \left.(
  \tfrac{1}{2} v_{1}^{2}-\tfrac{1}{2} v_{2}^{2}+h-\beta\eta
         )\right|_{x=x_{f}-0}
=-\beta\tfrac{1}{2}(U_{\infty}-U_{f})^{2}.
\end{equation}
If $v_{1}$, $v_{2}$, and $\eta$ at $x=x_{f}-0$ 
are eliminated from (\ref{5.640}) and (\ref{5.650}), we have 
\begin{equation}
  \label{5.660}
  (1-\beta)^{2}\tfrac{1}{2}(U_{\infty}-U_{f})^{2}
 =\left\{(1-\beta)+\tfrac{1}{2}\beta h_{b}\right\}h_{b}.
\end{equation}
Since $1-\beta$ is not very small, 
we can assume that $\tfrac{1}{2}\beta h_{b}\ll 1-\beta$; 
(\ref{5.660}) then yields 
\begin{equation}
  \label{5.670}
  U_{f}=U_{\infty}
 +\left(\frac{2h_{b}}{1-\beta}\right)^{\frac{1}{2}}.
\end{equation}

We can easily see that (\ref{5.670}) is 
the velocity formula obtained by von K\'arm\'an (1940), 
on the basis of Bernoulli's theorem, 
for a gravity current of the same kind. 
Hence, while von K\'arm\'an's argument leading to his formula 
was later shown to be invalid (Benjamin 1968), 
his formula itself is applicable 
when the lower vortex sheet is absent. 

When $\beta=\!\mbox{}+0$, 
von K\'arm\'an's velocity formula (\ref{5.670}) reduces to 
\begin{equation}
  \label{5.680}
  U_{f}=U_{\infty}+(2h_{b})^{\frac{1}{2}}.
\end{equation}
On the other hand, (\ref{5.680}) is also obtained 
from (\ref{5.590}) when $h_{b}\ll 1$. 
Here we recall that (\ref{5.590}) was derived 
from the front conditions based on 
the conservation laws of mass and of circulation. 
Thus we see that, when $\beta=\!\mbox{}+0$, 
i.e.\ when the Boussinesq approximation is adequate, 
von K\'arm\'an's formula can be derived 
on the basis of the conservation laws 
of mass and of circulation. 
This is a fact pointed out by Rotunno et al.\ (1988). 

\subsection{Gravity currents advancing along a no-slip upper boundary}
\subsubsection{Front conditions}
As in \S\,5.1.1, we consider the physical situation in \S\,3. 
However, in contrast to \S\,5.1.1, 
the upper-layer fluid is assumed to be advancing 
along the upper boundary as a gravity current. 
The front of the current is again identified 
with an inverted wall of fluid 
which is located at $x=x_{f}$ 
and is contained in the plane representing the frontal region. 

In \S\,5.1.1, the motion 
in the infinite interval $x\le x_{f}-0$ was considered 
to be specified in terms of the variables 
$u_{1}$, $u_{2}$, $h$, and $\eta$ introduced in \S\,3. 
However, it is preferable for our present purpose 
to employ, instead of $\eta$, the new variable $\zeta(x,t)$ 
which is so defined that $\rho_{1}\beta gH\zeta$ gives 
the pressure at the lower boundary for $|y|<\tfrac{1}{2}$. 
Thus we henceforth assume that the motion 
in the infinite interval $x\le x_{f}-0$ is specified 
in terms of $u_{1}$, $u_{2}$, $h$, and $\zeta$. 
The pressure distribution in this interval 
is expressed in terms of $\zeta$ by 
\begin{equation}
  \label{5.690}
  p_{\ast}=\left\{
  \begin{array}{lll}
    \rho_{1}\beta gH\zeta -\rho_{1}gHz,
                         & |y|<\tfrac{1}{2}, & 0 \le z \le h,\\[2pt]
    \rho_{1}\beta gH\zeta -\rho_{1}gHh -\rho_{2}gH(z-h),
                         & |y|<\tfrac{1}{2}, & h \le z \le 1.
  \end{array}
  \right.
\end{equation}

As for the motion in the infinite interval $x\ge x_{f}+0$, 
it can be specified in terms of $u_{1}$ and $\zeta$. 
Indeed, in this interval, the dimensionless velocity $u$ 
in (\ref{3.30}) is given by 
\begin{equation}
  \label{5.710}
  u=u_{1}(x,t),\qquad 0<z<1.
\end{equation}

Now, by an argument similar to that in \S\,5.1.1, 
we can find the front conditions. 
The one corresponding to the conservation law of circulation 
for the interfacial vortex sheet is expressed, 
in terms of the relative velocities (\ref{5.260}), by 
\begin{equation}
  \label{5.780}
  \left.\left\{
  \tfrac{1}{2} v_{1}^{2}-\tfrac{1}{2} v_{2}^{2}-\frac{(1-h)+\beta\zeta}{1-\beta}
         \right\}\right|_{x=x_{f}-0}
 +\left.\frac{\beta\zeta}{1-\beta}\right|_{x=x_{f}+0}
 +\left.
  \frac{\beta\tfrac{1}{2} v_{1}^{2}}{1-\beta}
  \right|_{x=x_{f}+0}
=-\tfrac{1}{2} U_{f}^{2}.
\end{equation}
Here the term on the right-hand side represents 
the influence of the upper vortex sheet. 
The remaining front conditions are as follows: 
\begin{equation}
  \label{5.790}
  \left.
  \begin{array}{c}
  \displaystyle
  \left.hv_{1}\right|_{x=x_{f}-0}
 -\left.v_{1}\right|_{x=x_{f}+0}=0,
  \quad
  \left.(1-\beta)(1-h)v_{2}\right|_{x=x_{f}-0}=0,\\[6pt]
  \displaystyle
  \left.\left\{
    hv_{1}^{2}+(1-\beta)(1-h)v_{2}^{2}+\tfrac{1}{2}(1-h)^{2}+\zeta
  \right\}\right|_{x=x_{f}-0}
 -\left.(v_{1}^{2}+\zeta)\right|_{x=x_{f}+0}=0.
  \end{array}
  \right\}
\end{equation}

\subsubsection{Energy condition and evolutionary condition}
Let us formulate the energy condition 
for the gravity current in \S\,5.3.1. 
In view of the first front condition 
in (\ref{5.790}), we can write 
\begin{equation}
  \label{5.800}
  \left.hv_{1}\right|_{x=x_{f}-0}
 =\left.v_{1}\right|_{x=x_{f}+0}
 =I_{f}(t).
\end{equation}
The energy condition is expressed by (\ref{5.310}) 
if we define $\Phi_{f}$ as follows: 
\begin{equation}
  \label{5.810}
  \Phi_{f}(t)
 =I_{f}
  \left\{
  \left.(\tfrac{1}{2} v_{1}^{2}+\zeta)\right|_{x=x_{f}-0}
 -\left.(\tfrac{1}{2} v_{1}^{2}+\zeta)\right|_{x=x_{f}+0}
  \right\}.
\end{equation}

Also, the evolutionary condition 
for the gravity current is expressed by (\ref{5.320}). 

\subsubsection{Velocity formula}
To find the velocity formula 
for the gravity current in \S\,5.3.1, we first put 
\begin{equation}
  \label{5.830}
  u_{1}=U_{\infty}, \quad\zeta=0, \qquad x\ge x_{f}+0.
\end{equation}

The velocity formula can now be derived 
from (\ref{5.780}) and (\ref{5.790}). 
Let us first consider, as in \S\,5.1.3, the case in which the density 
in the gravity current is almost equal to the ambient density, 
i.e.\ the case in which $\beta=\!\mbox{}+0$. 
In this case, (\ref{5.780}) reduces to 
\begin{equation}
  \label{5.840}
  \left.\left\{
  \tfrac{1}{2} v_{1}^{2}-\tfrac{1}{2} v_{2}^{2}-(1-h)
         \right\}\right|_{x=x_{f}-0}
=-\tfrac{1}{2} U_{f}^{2}.
\end{equation}
Using this together with (\ref{5.790}), 
we obtain the following formula: 
\begin{equation}
  \label{5.870}
  U_{f}
 =\frac{U_{\infty}}{1+h_{b}^{2}}
 +\frac{h_{b}
        \left\{
        2(1-h_{b})+2h_{b}^{2}(1-h_{b})-U_{\infty}^{2}
        \right\}^{\frac{1}{2}}}
       {1+h_{b}^{2}}.
\end{equation}

Comparing this formula with (\ref{5.380}) in \S\,5.1.3, 
we observe that (\ref{5.870}) is also obtained from (\ref{5.380}) 
when $h_{b}$ in (\ref{5.380}) is replaced by $1-h_{b}$. 
It follows from this fact that, when $\beta=\!\mbox{}+0$, 
the rate of advance of the gravity current is equal to 
that of a gravity current with the same dimensionless depth 
advancing along a no-slip lower boundary. 

We next consider the case in which 
the density in the gravity current is much smaller 
than the ambient density, i.e.\ the case in which $\beta=1-0$. 
In this case, (\ref{5.780}) yields 
\begin{equation}
  \label{5.880}
 -\left.\left\{(1-h)+\zeta\right\}\right|_{x=x_{f}-0}
 +\tfrac{1}{2}(U_{\infty}-U_{f})^{2}=0.
\end{equation}
From this and (\ref{5.790}), 
the velocity formula is obtained in the form 
\begin{equation}
  \label{5.900}
  U_{f}=U_{\infty}
 +\left\{
  \frac{h_{b}(1+h_{b})(1-h_{b})}{2-h_{b}}
  \right\}^{\frac{1}{2}}.
\end{equation}

In deriving (\ref{5.900}), we used (\ref{5.880}). 
While (\ref{5.880}) was obtained from (\ref{5.780}), 
it does not contain the term 
representing the influence of the upper vortex sheet. 
This implies that the presence of the upper vortex sheet 
is inessential for (\ref{5.900}) to be valid; 
(\ref{5.900}) applies 
even if the upper- and lower-layer fluids 
are allowed to slip at the upper boundary. 
This result is in accord with the fact 
that (\ref{5.900}) is the velocity formula 
derived by Benjamin (1968) for a gravity current 
of the same kind on the basis of inviscid-fluid theory. 

It can readily be shown, however, that (\ref{5.900}) 
is obtained only when the density in the gravity current 
is much smaller than the ambient density. 
We can therefore conclude as follows: within the framework of 
the one-dimensional two-layer shallow-water equations, 
Benjamin's formula 
applies only to a gravity current whose density 
is much smaller than the ambient density, 
i.e.\ a gravity current identifiable with a `cavity'. 

%% file: section6.tex
\section{Discussion}
A unified theory of internal bores and gravity currents 
has been developed within the framework 
of the one-dimensional two-layer shallow-water equations, 
and the validity of the theory has been confirmed 
in the light of some empirical facts. 
It should be borne in mind, however, 
that the theory is based on several assumptions. 
Hence the theory is not directly applicable 
to internal bores and gravity currents 
which violate the assumptions. 
In this section, we discuss two examples 
of such internal bores and gravity currents. 

\subsection{Internal hydraulic jumps}
An internal hydraulic jump is defined, in this paper, 
as a steady and stationary internal bore 
with the following property: 
the lower-layer fluid crosses the bore 
from the side on which the level of the interface is lower 
to the side on which it is higher. 
Our purpose is to examine how an internal hydraulic jump 
can be treated within the framework 
of the one-dimensional two-layer shallow-water equations. 

It may seem, at first sight, that an internal hydraulic jump 
can be treated as a shock satisfying 
the shock conditions (\ref{4.40})--(\ref{4.70}) 
with $U_{s}$ in them put equal to zero. 
We should note, however, that (\ref{4.40})--(\ref{4.70}) 
were derived on the assumption 
that the vortex sheet on a boundary 
never separates from the boundary. 
Hence (\ref{4.40})--(\ref{4.70}) lose their basis 
if this assumption is invalid. 
In fact, an internal hydraulic jump 
invalidates this assumption. 

To show this, we first note that 
the lower-layer fluid experiences 
an abrupt thickening of the lower layer 
when it crosses an internal hydraulic jump. 
Thus the lower-layer fluid is decelerated rapidly 
inside an internal hydraulic jump. 
However, it is well known that, 
when a steady flow along a rigid boundary 
is decelerated rapidly in the direction of the flow, 
the vortex sheet on the boundary separates from the boundary 
(see e.g.\ Batchelor 1967, \S\,5.10). 
Hence we can conclude that, inside an internal hydraulic jump, 
separation of the vortex sheet 
on the lower boundary occurs inevitably. 

We must take this separation into account 
in dealing with an internal hydraulic jump within the framework 
of the one-dimensional two-layer shallow-water equations. 
On the other hand, within the framework of the equations, 
separation of a vortex sheet cannot be treated directly. 
Hence, to resolve this difficulty, we introduce 
a model of an internal hydraulic jump 
in which the separation is represented in a indirect manner. 

In this model, it is supposed that the vortex sheet 
on the lower boundary is attached to the boundary 
everywhere inside the jump. 
Instead, to represent the separation of the vortex sheet, 
we assume the following process to take place 
somewhere inside the jump: 
the fluid in the vortex sheet leaks from the vortex sheet; 
it then rises through the lower layer 
and is absorbed into the interfacial vortex sheet. 
In compensation for this process, 
a fraction of the fluid out of the lower vortex sheet 
is absorbed into it. 

Now, within the framework of the one-dimensional 
two-layer shallow-water equations, 
this model jump can be represented by a shock. 
By applying the same argument as that in \S\,4.1, 
we can show that the shock satisfies, 
in place of (\ref{4.40}) with $U_{s}=0$, 
\begin{equation}
  \label{6.10}
  0=-\left[\tfrac{1}{2} u_{1}^{2}-\tfrac{1}{2} u_{2}^{2}+h-\beta\eta\right]
    +\int_{x_{s}-0}^{x_{s}+0}
     \frac{\left.\omega_{n\ast}\boldsymbol{u}_{\ast}\right|_{z=\mbox{}+0}}
          {\beta gH/L}
     \cdot\boldsymbol{k}\:\mathrm{d}x.
\end{equation}
The last term of this shock condition 
represents the advection of circulation 
due to the leakage of fluid from the lower vortex sheet, 
and the magnitude of the term is ${\it O}(1)$. 
We can also show that the remaining shock conditions 
are given by (\ref{4.50})--(\ref{4.70}) with $U_{s}=0$; 
the leakage of fluid from the lower vortex sheet 
does not affect these conditions. 

It is noteworthy that the shock satisfies 
(\ref{4.50}) and (\ref{4.60}) with $U_{s}=0$. 
These conditions yield the set of shock conditions (\ref{4.100}) 
with $U_{s}=0$ when the density in the upper layer is much smaller 
than that in the lower layer, i.e.\ when $\beta=1-0$. 
Thus, when $\beta=1-0$, an internal hydraulic jump can be treated 
as a shock satisfying (\ref{4.100}) with $U_{s}=0$; 
this result is, in fact, well known in classical hydraulics. 

On the other hand, when $\!\mbox{}+0\le\beta<1-0$, 
we must use (\ref{6.10}) to deal with an internal hydraulic jump. 
It is necessary then to express the last term of (\ref{6.10}) 
in terms of $u_{1}$, $u_{2}$, $h$, and $\eta$ 
at $x=x_{s}-0$ and at $x=x_{s}+0$. 
Unfortunately, however, 
there is no adequate means of doing this at present. 
As a result, the above treatment of an internal hydraulic jump 
is still incomplete when $\!\mbox{}+0\le\beta<1-0$. 

\subsection{Gravity currents of air 
            in a laboratory channel filled with water}
At the end of \S\,5.3.3, 
we discussed a gravity current whose density 
is much smaller than the ambient density, 
i.e.\ a gravity current identifiable with a `cavity'; 
it was found there that the velocity formula 
for the gravity current is given by (\ref{5.900}). 
Now let us return to the discussion 
of this specific kind of gravity current. 
Note that a gravity current of this kind is, in the following, 
referred to simply as a gravity current. 

This kind of gravity current 
was comprehensively studied by Benjamin (1968) 
on the basis of inviscid-fluid theory. 
In particular, he examined in great detail 
a gravity current free from dissipation of mechanical energy, 
i.e.\ a gravity current which satisfies 
\begin{equation}
  \label{6.20}
  \Phi_{f}=0,
\end{equation}
where $\Phi_{f}$ is defined by (\ref{5.810}). 
He showed that, as can be proved 
from (\ref{5.790}), (\ref{5.880}), and (\ref{6.20}), 
the depth and the rate of advance 
of such a gravity current are given by 
\begin{equation}
  \label{6.30}
  h_{b}=\tfrac{1}{2},\quad
  U_{f}=U_{\infty}+\tfrac{1}{2}.
\end{equation}
He also predicted that, if one end of a laboratory channel 
filled with water was opened, air would intrude 
into the channel as a gravity current free from dissipation. 
Gardner \& Crow (1970) and Wilkinson (1982) 
later confirmed this prediction by experiment. 

However, it can readily be verified 
that the evolutionary condition (\ref{5.320}) 
imposes the following restriction 
on the depth of a gravity current: 
\begin{equation}
  \label{6.40}
  0.653\le h_{b}\le 1-0.
\end{equation}
Apparently, a gravity current free from dissipation 
violates this restriction. 
This implies that such a gravity current 
does not satisfy (\ref{5.320}): 
it instead satisfies 
\begin{equation}
  \label{6.50}
  \lambda_{-}-U_{f}<\lambda_{+}-U_{f}<0
  \quad \mbox{at} \quad x=x_{f}-0.
\end{equation}
The present theory therefore leads us to the conclusion 
that a gravity current free from dissipation 
of mechanical energy cannot actually exist. 
It is evident that this conclusion disagrees with 
the experimental result of Gardner \& Crow and Wilkinson. 

It must be stressed here, however, 
that the present theory is based on the assumption 
that the thickness of the interfacial vortex sheet 
and the surface tension at the interface are negligible. 
As can be seen from the following argument, 
our theoretical conclusion is crucially 
dependent on this assumption. 

Let us now suppose that a gravity current 
free from dissipation is set up at an initial instant. 
This gravity current cannot remain free from dissipation 
if breaking of internal waves arises inside the frontal region. 
However, we observe from (\ref{6.50}) that 
there are no internal waves coming into the region from behind. 
It seems, therefore, that breaking of internal waves 
does not arise inside the frontal region. 

However, we should realize that, 
if the gravity current satisfies the assumption stated above, 
Kelvin-Helmholtz instability can arise inside the frontal region. 
Indeed, it is seen from the theory of 
Kelvin-Helmholtz instability (see e.g.\ Drazin \& Reid 1981, \S\,4) 
that, if the gravity current satisfies the assumption, 
the interfacial vortex sheet is unstable 
to internal waves whose lengths are 
sufficiently smaller than the length 
\begin{equation}
  \label{6.60}
  \xi_{g}
 =\frac{\rho_{1}\rho_{2}\Upsilon^{2}}
       {(\rho_{1}^{2}-\rho_{2}^{2})g}
 =\frac{(1-\beta)\Upsilon^{2}}{(2-\beta)\beta g},
\end{equation}
where $\Upsilon$ is the scale of 
the dimensional velocity difference across the vortex sheet. 

Accordingly, if the gravity current satisfies 
the assumption stated above, 
it is inferred that the following process 
takes place inside the frontal region: 
short internal waves are generated and grow 
through Kelvin-Helmholtz instability; 
these waves eventually break 
to cause dissipation of mechanical energy. 
This explains why the present theory leads to the conclusion 
that a gravity current free from dissipation is impossible. 

Having seen the reason for 
the impossibility in the present theory 
of a gravity current free from dissipation, 
we next examine what conditions are necessary 
for our theoretical conclusion to be right 
when the thickness of the interfacial vortex sheet 
and the surface tension at the interface 
are not necessarily negligible. 

We consider again a gravity current 
free from dissipation which is set up at an initial instant, 
and direct our attention to the interfacial vortex sheet. 
As explained above, the gravity current cannot remain 
free from dissipation if the vortex sheet 
is unstable inside the frontal region. 
If we ignore the thickness of the vortex sheet 
and the surface tension at the interface, 
we can again expect that the vortex sheet is unstable 
to internal waves whose lengths are sufficiently smaller 
than the length $\xi_{g}$ defined by (\ref{6.60}). 

However, the vortex sheet is, in fact, 
stable to internal waves with lengths 
sufficiently small in comparison with its thickness 
(see e.g.\ Drazin \& Reid 1981, \S\,23). 
Thus, for the vortex sheet to be 
unstable inside the frontal region, 
the gravity current must satisfy 
\begin{equation}
  \label{6.70}
  J=\frac{\delta_{f}}{\xi_{g}}
   =\delta_{f}\frac{(\rho_{1}^{2}-\rho_{2}^{2})g}
                   {\rho_{1}\rho_{2}\Upsilon^{2}}
   =\delta_{f}\frac{(2-\beta)\beta g}{(1-\beta)\Upsilon^{2}}\ll 1,
\end{equation}
where $\delta_{f}$ is the scale of the thickness 
of the vortex sheet inside the frontal region. 

The vortex sheet is also stable to 
internal waves with lengths 
sufficiently smaller than the following length $\xi_{t}$ 
(see e.g.\ Drazin \& Reid 1981, p.~28): 
\begin{equation}
  \label{6.80}
  \xi_{t}
 =\frac{\gamma}{(\rho_{1}-\rho_{2})g}
  \frac{(\rho_{1}^{2}-\rho_{2}^{2})g}
       {\rho_{1}\rho_{2}\Upsilon^{2}}
 =\frac{\gamma}{\rho_{1}\beta g}
  \frac{(2-\beta)\beta g}{(1-\beta)\Upsilon^{2}},
\end{equation}
where $\gamma$ is the surface tension at the interface. 
Thus, for the vortex sheet to be 
unstable inside the frontal region, 
the following condition must also be satisfied: 
\begin{equation}
  \label{6.90}
  T=\frac{\xi_{t}}{\xi_{g}}
   =\frac{\gamma}{(\rho_{1}-\rho_{2})g}
    \left\{\frac{(\rho_{1}^{2}-\rho_{2}^{2})g}
                {\rho_{1}\rho_{2}\Upsilon^{2}}\right\}^{2}
   =\frac{\gamma}{\rho_{1}\beta g}
    \left\{\frac{(2-\beta)\beta g}{(1-\beta)\Upsilon^{2}}\right\}^{2}
    \ll 1.
\end{equation}

Accordingly, we see that our theoretical conclusion 
applies to the gravity current only when 
it satisfies both the conditions (\ref{6.70}) and (\ref{6.90}); 
otherwise dissipation of mechanical energy 
does not arise inside the frontal region 
contrary to our theoretical conclusion. 

We are now in a position to explain why our theoretical conclusion 
disagrees with the experimental result 
of Gardner \& Crow and Wilkinson. 
They found that air can intrude into a laboratory channel 
filled with water as a gravity current free from dissipation. 
In view of the above result, we can expect 
that such a gravity current violates one or both 
of the conditions (\ref{6.70}) and (\ref{6.90}); 
we wish to confirm this in the following. 

Let us consider a gravity current of air advancing 
without dissipation into water in a laboratory channel. 
For this gravity current, we may put 
\begin{equation}
  \label{6.100}
  \Upsilon=(\beta gH)^{\frac{1}{2}},
\end{equation}
where $H$ has been used again to denote the depth of the channel. 
We may also consider that $\delta_{f}$ is given by 
the following formula (see e.g.\ Batchelor 1967, \S\,5.12): 
\begin{equation}
  \label{6.110}
  \delta_{f}=(\nu_{a}l_{f}/\Upsilon)^{\frac{1}{2}}.
\end{equation}
Here $\nu_{a}$ is the kinematic viscosity of air, 
which has been used rather than the kinematic viscosity 
of water $\nu_{w}$ because $\nu_{a}>\nu_{w}$; 
and $l_{f}$ is the length scale of the frontal region. 
It is reasonable, however, to assume that 
\begin{equation}
  \label{6.120}
  l_{f}/H={\it O}(1).
\end{equation}
Thus the value of $\delta_{f}$ may be estimated 
from (\ref{6.110}) with $l_{f}$ in it replaced by $H$. 

Now, to estimate the values of $J$ and $T$ 
for the gravity current, we put 
\begin{equation}
  \label{6.130}
  \left.
  \begin{array}{c}
  \displaystyle
  \rho_{1}=1.0\times 10^{3}{\mbox{$\;$kg\,m$^{-3}$}},\quad
  \rho_{2}=1.2{\mbox{$\;$kg\,m$^{-3}$}},\quad
  g=9.8{\mbox{$\;$m\,s$^{-2}$}}\\[2pt]
  \displaystyle
  \nu_{a}=1.5\times 10^{-5}{\mbox{$\;$m$^{2}$\,s$^{-1}$}},\quad
  \gamma=7.3\times 10^{-2}{\mbox{$\;$N\,m$^{-1}$}},\quad
  H=4.0\times 10^{-1}{\mbox{$\;$m}},
  \end{array}
  \right\}
\end{equation}
where we have assumed that $H$ is equal to the depth 
of the deepest channel used in the experiments of Wilkinson. 
Using these values, we obtain 
\begin{equation}
  \label{6.140}
  J=3.6,\quad T=3.2\times 10.
\end{equation}
This shows that, as expected, the gravity current 
satisfies neither (\ref{6.70}) nor (\ref{6.90}). 
Hence it is quite natural for our theoretical conclusion 
not to apply to the gravity current. 

Finally, we should note the following fact: 
when gravity currents free from dissipation can exist, 
those not free from dissipation 
must have $h_{b}$ lying in the range 
\begin{equation}
  \label{6.150}
  0.781\le h_{b}\le 1-0.
\end{equation}
This was demonstrated by Wilkinson (1982) 
both experimentally and theoretically. 
The implication of this fact is that, 
when gravity currents free from dissipation can exist, 
no gravity current can advance faster than them. 
Hence, in a laboratory channel (of normal depth) filled with water, 
no gravity current of air 
can have $U_{f}$ larger than $U_{\infty}+\tfrac{1}{2}$.

%% file: section7.tex
\section{Conclusion}
Internal bores and gravity currents in a two-fluid system 
can be treated, in a unified way, on the basis of 
the four basic laws of 
the one-dimensional two-layer shallow-water equations: 
the conservation laws of mass for the upper layer, 
of mass for the lower layer, of momentum for the layers together, 
and of circulation for the interfacial vortex sheet. 

%% file: appendix.tex
\appendix
\section*{Appendix.\hspace{0.5em}Derivation of (\ref{4.240})}
\addcontentsline{toc}{section}{Appendix}
\addtocounter{section}{1}
In this appendix, lower-case Latin indices are used 
to represent the numbers 1, 2, and 3; 
in addition, the summation convention is implied. 

Let us first define a function $u_{3}(x,t)$ by $u_{3}=h$. 
Then (\ref{4.200}) can be written as 
\begin{equation}
  \label{A.10}
  \Theta_{ij}\frac{\partial u_{j}}{\partial t}
 +\vartheta_{ij}\frac{\partial u_{j}}{\partial x}=0,
\end{equation}
where $\Theta_{ij}$ and $\vartheta_{ij}$ are 
the $(i,j)$ components of the matrices 
\begin{equation}
  \label{A.20}
  (\Theta_{ij})
 =
  \left(
  \begin{array}{ccc}
  0&0&0\\
  0&0&1\\
  1&-(1-\beta)&0
  \end{array}
  \right),
  \quad
  (\vartheta_{ij})
 =
  \left(
  \begin{array}{ccc}
  h&1-h&u_{1}-u_{2}\\
  h&0&u_{1}\\
  u_{1}&-(1-\beta)u_{2}&1
  \end{array}
  \right).
\end{equation}
The characteristic velocities $\lambda_{+}$ and $\lambda_{-}$ 
are given by the finite real roots of the following equation 
for an unknown $\lambda$ (see e.g.\ Whitham 1974, \S\,5.1): 
\begin{equation}
  \label{A.30}
  \mbox{det}(\lambda\Theta_{ij}-\vartheta_{ij})=0.
\end{equation}
Now let $\hat{\lambda}$ denote $\lambda-U_{s}$. 
Then, in terms of $\hat{\lambda}$, 
(\ref{A.30}) can be rewritten as follows: 
\begin{equation}
  \label{A.40}
  \mbox{det}(\hat{\lambda}\Theta_{ij}-\hat{\vartheta}_{ij})=0.
\end{equation}
Here $\hat{\vartheta}_{ij}=\vartheta_{ij}-U_{s}\Theta_{ij}$; 
using the relative velocities (\ref{4.80}), we have 
\begin{equation}
  \label{A.50}
  (\hat{\vartheta}_{ij})
 =
  \left(
  \begin{array}{ccc}
  h&1-h&v_{1}-v_{2}\\
  h&0&v_{1}\\
  v_{1}&-(1-\beta)v_{2}&1
  \end{array}
  \right).
\end{equation}
Thus we can obtain $\lambda_{+}-U_{s}$ and $\lambda_{-}-U_{s}$ 
by solving (\ref{A.40}) for $\hat{\lambda}$. 
It can readily be shown that (\ref{A.40}) 
yields the following quadratic equation for $\hat{\lambda}$: 
\begin{equation}
  \label{A.60}
  \hat{\lambda}^{2}
 -\frac{(1-h)v_{1}+(1-\beta)hv_{2}}{(1-\beta)h+(1-h)}2\hat{\lambda}
 +\frac{(1-h)v_{1}^{2}+(1-\beta)hv_{2}^{2}-h(1-h)}{(1-\beta)h+(1-h)}
 =0.
\end{equation}
Hence, solving this quadratic equation, we find (\ref{4.240}).